\pdfoutput=1
\documentclass[preprint,authoryear]{elsarticle}

\usepackage{amssymb}

\usepackage{amsmath}

\usepackage{makecell}

\usepackage{lineno} 

\usepackage{hyperref}

\usepackage{subcaption}

\usepackage{tikz}

\usepackage{multirow}

\usepackage{float}

\usetikzlibrary{arrows.meta,positioning}

\journal{Ocean \& Coastal Management}

\begin{document}

\begin{frontmatter}

\title{\textit{ShipEcho} - An Interactive Tool for Global Mapping of Underwater Radiated Noise from Vessels}

\author[1]{Mark Shipton}
\author[3]{Valentino Denona}
\author[3,2]{Đula Nađ}
\author[1,2,3]{Roee Diamant\corref{cor1}}

\cortext[cor1]{Corresponding author}
\ead{roee.d@univ.haifa.ac.il}

\affiliation[1]{organization={University of Haifa, Charney School of Marine Sciences},
                addressline={199 Aba Khoushy Ave. Mount Carmel},
                city={Haifa},
                postcode={3498838},
                state={},
                country={Israel}}

\affiliation[2]{organization={University of Zagreb, Faculty of Electrical Engineering and Computing},
                addressline={Unska 3},
                city={Zagreb},
                postcode={HR-10000},
                state={},
                country={Croatia}}

\affiliation[3]{organization={CoE MARBLE - Centre of Excellence in Maritime Robotics and Technologies for Sustainable Blue Economy},
                addressline={Unska 3},
                city={Zagreb},
                postcode={HR-10000},
                state={},
                country={Croatia}}

\begin{abstract}
Underwater radiated noise from vessels (V-URN) is a recognized environmental stressor that negatively impacts marine ecosystems. Significant resources are invested in the development of V-URN monitoring indicators, regulatory frameworks, and management-oriented assessments. One approach with high potential for impact is V-URN mapping, which can provide actionable spatiotemporal information for environmental assessment and mitigation planning. Producing management-scale maps remains challenging as passive acoustic measurements are spatially sparse and many operational systems depend on specialist workflows and costly access to wide-area vessel activity data. To address these constraints, we introduce \textit{ShipEcho}, a freely accessible web-based Geographic Information System (GIS) that provides near-real-time V-URN mapping using vessel data acquired through a community-based AIS exchange. Using established vessel SL models and propagation modeling informed by bathymetric data, \textit{ShipEcho} produces near-real-time and cumulative noise maps across regions worldwide. These include sound pressure levels and sound exposure levels using standard indicators, including the 63~Hz and 125~Hz one-third octave bands and a 20--2000~Hz broadband level. We describe the system architecture, data pipeline, modeling workflow, and key assumptions, and evaluate map accuracy through comparison with acoustic recordings. We then demonstrate how \textit{ShipEcho} can support management-level assessment, decision-making, and policy initiatives through practical use cases.
\end{abstract}

\begin{graphicalabstract}
\centering
\includegraphics[scale=0.35]{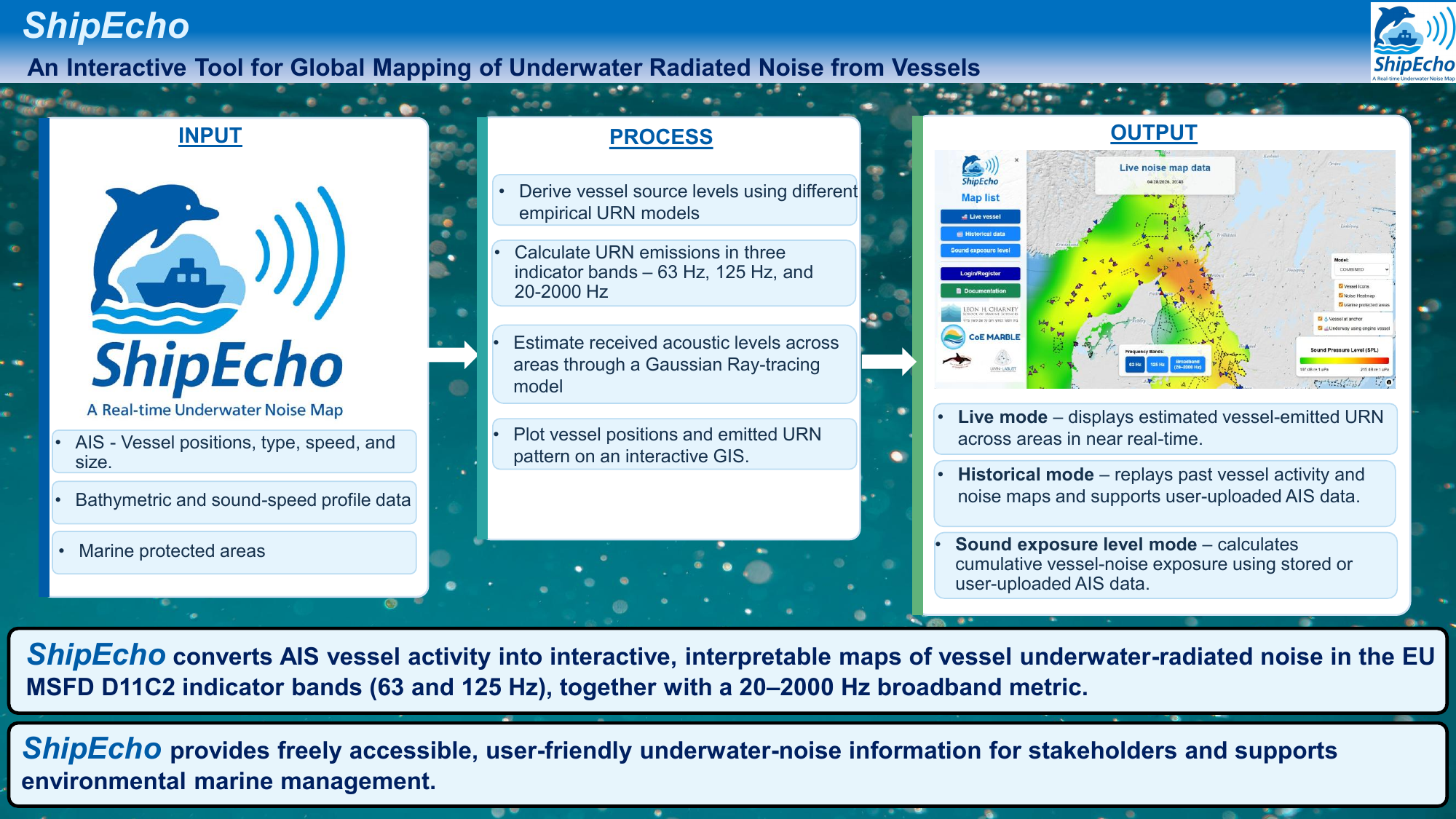}
\end{graphicalabstract}

\begin{highlights}
\item \textit{ShipEcho} provides near-real-time and cumulative maps of vessel underwater radiated noise.
\item The platform combines AIS traffic data, source-level models, environmental layers, and acoustic propagation modeling.
\item Outputs include SPL and SEL in the 63~Hz and 125~Hz one-third-octave bands and a 20--2000~Hz broadband metric.
\item Users can compare multiple source-level models in an interactive web-based GIS environment.
\item The system supports management-oriented screening of vessel-noise exposure and mitigation scenarios.
\end{highlights}


\end{frontmatter}



\section{Introduction}
\label{intro}
Vessel underwater radiated noise (V-URN) is increasingly recognized as a pervasive, chronic pressure on marine ecosystems, driven in large part by commercial shipping that introduces substantial and persistent low-frequency energy overlapping biologically important signals~\cite{pine2016potential,chapman2011_lowfreq_trend,mcdonald2006_increases_noise}. Prolonged elevation of noise levels can reduce effective communication and sensing ranges via masking, modify movement and foraging behavior, and contribute to physiological stress across taxa~\cite{erbe2019_shipnoise_review,slabbekoorn2010_noisy_spring}. These documented effects motivate sustained monitoring and mapping of V-URN to support risk assessment, trend evaluation, and the design and appraisal of mitigation measures.

Empirical characterizations of V-URN are commonly derived from passive acoustic monitoring (PAM), where fixed recorders provide high-quality measurements but remain spatially sparse and are typically analyzed offline. As a result, PAM-only assessments are hard to scale to the spatial extents and temporal continuity implied by management considerations. Furthermore, long-term offshore deployments face practical constraints (instrument reliability, servicing logistics) that limit sustained coverage. To overcome this, current practice increasingly combines measurements with numerical modeling to produce statistical underwater noise maps. For example, the Joint Monitoring Program for Ambient Noise North Sea (JOMOPANS) integrates vessel data and a source level (SL) estimation model with environmental data and localized measurements to construct underwater noise map reports over the area of the North Sea~\cite{dejong2022_north_sea_sound_maps_2019_2020}. Nevertheless, a gap persists between rigorous, expert-driven mapping workflows and the needs of regulatory authorities and stakeholders for transparent, interactive, and easily accessible V-URN information at decision-maker scales. Existing implementations require specialist configuration, substantial environmental inputs, and continuous vessel-traffic data. In particular, real-time AIS access at large spatial scales and over long durations is often obtained via commercial providers, and the associated costs can become a barrier to developing, sustaining, and routinely updating large-area noise-mapping systems. Moreover, current mapping products are disseminated as static reports~\cite{dejong2022_north_sea_sound_maps_2019_2020} or restricted-access platforms~\cite{quonops_online}, which limit exploratory use and reproducibility.

Methodologically, V-URN mapping depends on SL models that infer spectral levels from AIS-accessible predictors such as vessel length, speed, and type~\cite{breeding1996randi,macgillivray2021reference}. Yet many existing SL models apply only to subsets of vessel classes~\cite{chion2019_meta_analysis_ship_source_levels}, and most applications adopt only a single default SL model. For example, the JOMOPANS North Sea noise maps apply only the JOMOPANS-ECHO model~\cite{macgillivray2021reference}, constraining cross-model sensitivity assessment. Converting SL to received levels further requires propagation modeling that is sensitive to bathymetry, seabed properties, sound-speed structure, and sea state, while high-fidelity approaches such as parabolic equations and normal modes can be computationally burdensome when repeatedly evaluated over large grids and frequent update cycles~\cite{West1992_PEmodelTutorial,graves1974normal_mode_calculations}.

This paper presents the architecture and modeling workflow of \textit{ShipEcho}, a web-based Geographic Information System (GIS) for V-URN mapping. \textit{ShipEcho} is designed to address existing practical and methodological gaps by providing a freely accessible, interactive GIS tool that integrates vessel activity data from AIS and environmental data from open sources in a transparent, exploratory interface. To reduce a key operational cost barrier, \textit{ShipEcho} relies on community-shared AIS provision (\textit{AISHub})~\cite{aishub} rather than proprietary AIS services. To avoid reliance on any single SL assumption set and to broaden applicability across vessel types, \textit{ShipEcho} implements five SL models designed for different vessel types and allows users to select and compare them, enabling direct examination of how SL-model choice affects mapped V-URN distributions. For propagation, \textit{ShipEcho} employs a computationally efficient Gaussian ray-tracing approach that accounts for bathymetry and sound-speed profile (SSP), providing a pragmatic balance between physical realism and computational cost suited to interactive mapping~\cite{porter1987_gaussian_beam_tracing}. Finally, to keep outputs directly interpretable in management settings, \textit{ShipEcho} reports sound pressure levels (SPL) and sound exposure levels (SEL) in commonly adopted indicator bands, including the 63~Hz and 125~Hz one-third-octave bands used under the MSFD~\cite{ec2017_848_msfd_descriptor11}, alongside a 20--2000~Hz broadband metric used in applied marine soundscape and vessel-noise studies~\cite{vieira2021meagre}. 

Beyond providing instantaneous visualization, \textit{ShipEcho} is intended as a practical monitoring aid. It enables users to repeatedly query and map V-URN over consistent indicator bands, compare conditions across dates and seasons, and rapidly screen for spatio-temporal V-URN ``hotspots'' associated with dense vessel activity. The platform is designed to be transferable across regions with AIS coverage and baseline environmental layers, allowing the same workflow to be applied across jurisdictions without re-engineering the mapping pipeline. To support ecologically focused interpretation, \textit{ShipEcho} overlays marine protected areas (MPAs) directly on the map, allowing users to contextualize predicted SPL and SEL patterns relative to protected features and to extract summaries within designated areas for reporting and comparison. In addition, users can upload AIS datasets locally, allowing scenario-based analysis, such as evaluating the effect of vessel-speed limits on V-URN SEL.
The following sections describe the \textit{ShipEcho} architecture and modeling workflow, including data inputs, user interaction, SL modeling, propagation modeling, and GIS-based visualization of V-URN distributions. We summarize the main assumptions and limitations of the web-based GIS framework and outline its intended decision-support applications for managers and other stakeholders. Finally, we present a representative example showing how \textit{ShipEcho} can be used for underwater-noise monitoring and management-oriented scenario assessment. \textit{ShipEcho} can be freely accessed at \url{https://noisemaps.marble.eu/}.

\section{Methods}
\label{sec:Methods}
\subsection{System Overview and Functionality}
\label{sec:Functionality}
\textit{ShipEcho} is an interactive web-based GIS for visualizing V-URN. It provides three display modes: Live vessel mode (LVM), historical mode (HM), and sound exposure level mode (SELM). LVM displays near-real-time V-URN estimates based on the latest AIS updates. HM enables replay of vessel activity for user-selected dates. SELM summarizes cumulative exposure by visualizing the spatial distribution of sound exposure level (SEL) over the mapped domain. A detailed description of each mode is provided in~\ref{app1}.

\begin{figure}[!htbp]
\centering
\scriptsize
\resizebox{0.98\linewidth}{!}{%
\begin{tikzpicture}[
  node distance=5mm and 6mm,
  line/.style={-Latex, thick},
  block/.style={
    draw,
    rounded corners,
    align=center,
    text width=31mm,
    minimum height=10mm,
    inner sep=4pt
  }
]

\node[block] (ais) {AIS feed\\(AISHub)};
\node[block, right=of ais] (ingest) {AIS preprocessing\\decode, filter, store};
\node[block, right=of ingest] (attr) {Vessel attributes\\MMSI, category, length, speed, draft, position};
\node[block, right=of attr] (sl) {SL estimation\\specific SL model or Combined};

\node[block, below=of ais] (env) {Environmental inputs\\bathymetry and SSP};
\node[block, right=of env] (prop) {Propagation\\Gaussian ray tracing\\TL over grid};
\node[block, right=of prop] (map) {Mapped levels\\$\mathrm{RL}=\mathrm{SL}-\mathrm{TL}$\\SPL and SEL bands};
\node[block, right=of map] (gis) {Web GIS visualization\\heatmap, vessels, MPA overlay};

\draw[line] (ais) -- (ingest);
\draw[line] (ingest) -- (attr);
\draw[line] (attr) -- (sl);

\draw[line] (sl.south) |- (prop.north);
\draw[line] (env) -- (prop);
\draw[line] (prop) -- (map);
\draw[line] (map) -- (gis);

\end{tikzpicture}%
}
\caption{Processing workflow implemented in \textit{ShipEcho}. AIS-derived vessel attributes are used to estimate SL using selectable models, while environmental inputs support propagation modeling and TL estimation. The resulting SPL and SEL fields are mapped in the selected indicator bands and visualized in the web GIS together with vessel symbols and optional MPA overlays.}
\label{fig:shipecho_pipeline}
\end{figure}
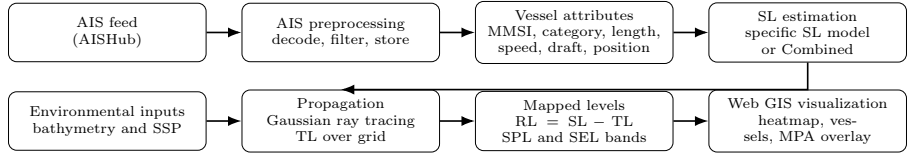

Across modes, \textit{ShipEcho} follows a common end-to-end processing workflow as described in Fig.~\ref{fig:shipecho_pipeline}. The workflow begins with the acquisition of AIS messages via \textit{AISHub}. The public implementation of \textit{ShipEcho} queries AIS across broad areas, but persistent storage and downstream SEL computations are restricted to predefined maritime regions, excluding inland subregions, to maintain update times compatible with interactive operation\footnote{To maintain interactive computation times, the current \textit{ShipEcho} implementation restricts persistent AIS storage and SEL calculations to predefined regions. The supported regions are listed in the project repository.} AIS decoding and preprocessing then generate a consistent stream of time-stamped vessel states. For each AIS report, \textit{ShipEcho} identifies the vessel and extracts the attributes required for noise prediction (e.g., vessel category, length, speed, draft, and position). These inputs are passed to the user-selected SL model to estimate vessel acoustic output in the selected indicator band. Predicted SL is then combined with environmental inputs, primarily bathymetry and the sound-speed profile (SSP), within a computationally efficient Gaussian ray-tracing propagation model to estimate transmission loss (TL) over the mapped grid. Finally, received levels are computed and aggregated on the spatial grid to generate map layers reporting SPL and SEL in the selected band. The resulting layers are served as map tiles and updated to support interactive exploration in LVM, HM, and SELM.

\begin{figure}[!htbp]
\centering
\includegraphics[width=0.9\textwidth]{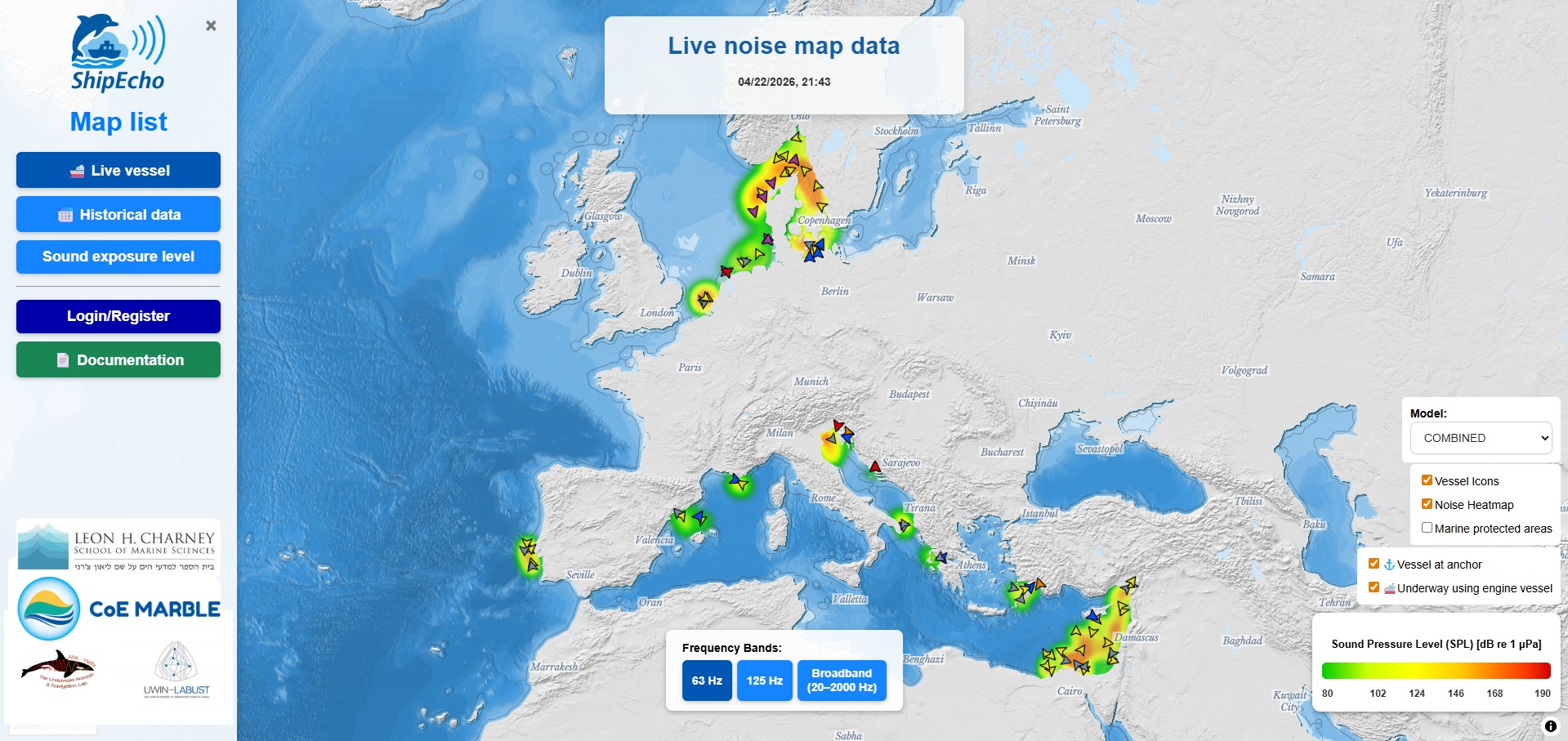}
\caption{Common user-interface overlays in \textit{ShipEcho} (example view of 'Live vessel mode'), including the basemap, vessel symbols, band-selection controls, the V-URN heatmap layer, and layer-selection options.}
\label{fig:ui_overlays}
\end{figure}

All modes provide a consistent set of map overlays and controls, illustrated in Fig.~\ref{fig:ui_overlays}, that allow users to adjust processing options and display layers. In all modes, the interactive map is the primary interface element, enabling users to pan and zoom to areas of interest. Ocean areas are rendered in blue tones with bathymetry shading to provide depth context, while land is shown with higher contrast, and major cities are labeled for geographic orientation. Estimated V-URN levels are visualized as a color-coded heatmap with an associated scale. Upon user selection, an overlay of legally designated marine protected areas (MPAs) from the World Database on Protected Areas (WDPA) can be displayed~\cite{UNEPWCMC_IUCN_WDPA_2026}. Vessels are shown as triangular symbols, with symbol color indicating vessel category. To maintain responsiveness over large spatial extents, vessel symbols are aggregated at low zoom levels and progressively revealed as users zoom in.

\begin{figure}[!htbp]
\centering
\includegraphics[width=0.8\textwidth]{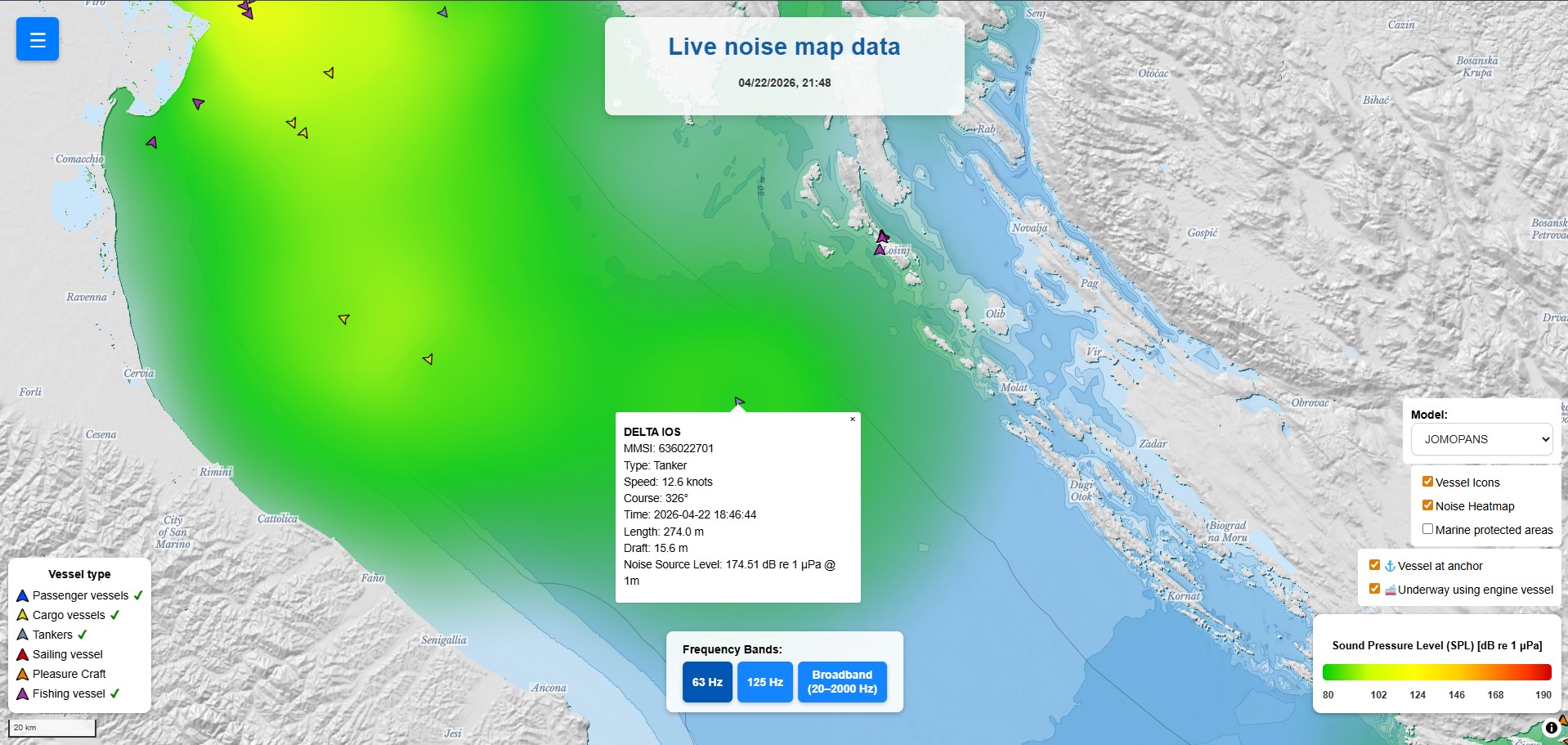}
\caption{Example of the vessel-information panel displayed when a vessel is selected in \textit{ShipEcho}.}
\label{fig:vessel_select}
\end{figure}

Users select the output band at the bottom of the display (63~Hz, 125~Hz, or 20--2000~Hz broadband). A control panel allows users to choose the vessel SL model (Section~\ref{subsec:sl_models}) and to toggle layers, including vessel symbols, the V-URN heatmap, MPAs, and vessel subsets based on AIS navigation status (for example, underway or at anchor). As shown in Fig.~\ref{fig:vessel_select}, selecting a vessel opens an information panel reporting vessel metadata and recent AIS state, including name, last update time, Maritime Mobile Service Identity (MMSI), speed, course, length, draft, and the estimated SL under the currently selected SL model.

\subsection{V-URN Modeling}
\label{sec:V-URN_Modeling}

\subsubsection{Vessel Source-Level Models}
\label{subsec:sl_models}
Vessel SL models provide a practical means of estimating V-URN when direct source measurements are unavailable. Modeling is performed by predicting spectral levels based on data obtained via ship registries and AIS (e.g., vessel type, length, and speed). Parametric formulations such as the RANDI 3.1 model~\cite{breeding1996randi} are widely used because they are computationally efficient and therefore suitable for large-scale, repeated evaluation in noise-mapping applications. In these models, the radiated spectrum is represented by frequency-dependent coefficients that vary by vessel class and are combined with scaling terms that capture the dominant dependence of radiated noise on vessel size and speed, yielding band or spectral estimates that can be used in tandem with propagation models to predict received levels across areas. A practical consideration is that SL models differ in their underlying calibration datasets and intended scope; consequently, not all formulations provide coefficients or parameterizations for every vessel category. 

In \textit{ShipEcho}, vessel activity is represented by AIS messages indexed by the vessel's Maritime Mobile Service Identity (MMSI) identifier and timestamp. To support V-URN mapping across heterogeneous traffic and to enable exploration of model-dependent variability,~\textit{ShipEcho} implements five SL models that can be selected by users: (1) the Research Ambient Noise Directionality (RANDI) 3.1 model~\cite{breeding1996randi}, (2) the JOMOPANS-ECHO (JE) model~\cite{macgillivray2021reference}, (3) the Length--Breadth--Draft--Speed (LBDS) model~\cite{simard2016analysis}, (4) the Achieve Quieter Oceans by shipping noise footprint reduction (AQUO) model~\cite{Audoly2015AQUO_R2_9}, and (5) our model for small recreational vessels (SRV)~\cite{Shipton2026SmallVesselURN}. A detailed description of the SL models used in \textit{ShipEcho} is provided in~\ref{app2}. In addition, \textit{ShipEcho} provides a \textit{Combined} model option that reports the energetic mean of the SL estimates from all available models for a given vessel in each indicator band. The \textit{Combined} model is calculated by
\begin{equation}
L_{\mathrm{Combined},b}
=
10\log_{10}\!\left(
\frac{1}{M}\sum_{m=1}^{M} 10^{L_{m,b}/10}
\right),
\quad
b \in \{63,\,125,\,20\text{--}2000~\mathrm{Hz}\}.
\label{eq:combined_energetic_mean}
\end{equation}
where \(L_{m,b}\) denotes the SL estimate produced by model \(m\) in band \(b\), and \(M\) is the number of models included in the combination. This multi-model capability allows users to select among alternative SL formulations. The vessel type reported in AIS is used by the selected SL model to parameterize the SL prediction, enabling direct comparison of mapped V-URN patterns under different source-model assumptions. The vessel-type coverage of each SL model is summarized in Table~\ref{tab:model-ais-types}.

\begin{table}[htbp]
\footnotesize
\centering
\setlength{\tabcolsep}{4pt}
\caption{Vessel-type coverage of the URN estimation models implemented in \textit{ShipEcho}. AIS type codes are defined as follows: 60--69 denote passenger vessels, including 60 (passenger), 61--64 (passenger carrying hazardous-category A--D cargo), 65--68 (reserved), and 69 (passenger, no additional information); 70--79 denote cargo vessels, including 70 (cargo), 71--74 (cargo carrying hazardous-category A--D cargo), 75--78 (reserved), and 79 (cargo, no additional information); 80--89 denote tankers, including 80 (tanker), 81--84 (tanker carrying hazardous-category A--D cargo), 85--88 (reserved), and 89 (tanker, no additional information); 30 denotes fishing vessels, 36 denotes sailing vessels, and 37 denotes pleasure craft.}
\label{tab:model-ais-types}
\begin{tabular}{lcccccc}
\hline
\textbf{Model} &
\makecell[c]{\textbf{Cargo}\\(AIS 70--79)} &
\makecell[c]{\textbf{Tankers}\\(AIS 80--89)} &
\makecell[c]{\textbf{Passenger}\\(AIS 60--69)} &
\makecell[c]{\textbf{Fishing}\\(AIS 30)} &
\makecell[c]{\textbf{Pleasure}\\(AIS 37)} &
\makecell[c]{\textbf{Sailing}\\(AIS 36)} \\
\hline
RANDI & \checkmark & \checkmark & \checkmark &            &            &            \\
JE    & \checkmark & \checkmark & \checkmark & \checkmark &            &            \\
LBDS  & \checkmark & \checkmark & \checkmark &            &            &            \\
AQUO  & \checkmark & \checkmark & \checkmark & \checkmark & \checkmark & \checkmark \\
SRV   &            &            &            &            & \checkmark & \checkmark \\
\hline
\end{tabular}
\end{table}

All models except LBDS provide an estimated source spectral density level at 1~Hz resolution, i.e., a source-level PSD, \(L_{\mathrm{PSD}}(f)\), in units of dB re \(1~\mu\mathrm{Pa}^2/\mathrm{Hz}\) referenced to 1~m. In contrast, LBDS provides an SL estimate directly in standard one-third-octave bands. To reduce computational cost while preserving accurate band-integrated energy, the integral of the linear PSD within each one-third-octave band, including the 63~Hz and 125~Hz indicator bands and the bands used to form the 20--2000~Hz broadband indicator, is approximated using the five-point Boole's rule~\cite{weisstein_boolesrule}. The modeled spectral density level is first converted to linear units according to
\begin{equation}
\label{eq:psd_linear}
S_p(f) = 10^{L_{\mathrm{PSD}}(f)/10},
\qquad
\left[\mu\mathrm{Pa}^2/\mathrm{Hz}\right].
\end{equation}
For a one-third-octave band defined by its center frequency \(f_c\), the lower and upper band-edge frequencies are given by
\begin{equation}
\label{eq:band_edges}
f_\mathrm{L} = f_c\,2^{-1/6},
\qquad
f_\mathrm{U} = f_c\,2^{1/6},
\qquad
\Delta f = f_\mathrm{U}-f_\mathrm{L}.
\end{equation}
Boole's rule requires equally spaced evaluation points over the integration interval \([f_\mathrm{L},f_\mathrm{U}]\). Defining the step size as \(h=\Delta f/4\), the PSD is evaluated at the five quadrature frequencies
\begin{equation}
\label{eq:boole_points}
\begin{cases}
f^{(0)} = f_\mathrm{L},\\[4pt]
f^{(1)} = f_\mathrm{L} + h,\\[4pt]
f^{(2)} = f_\mathrm{L} + 2h,\\[4pt]
f^{(3)} = f_\mathrm{L} + 3h,\\[4pt]
f^{(4)} = f_\mathrm{U},
\end{cases}
\qquad
h=\dfrac{\Delta f}{4}.
\end{equation}
The band-integrated mean-square pressure, i.e., the band energy in linear units, is then approximated by
\begin{equation}
\label{eq:band_energy}
\begin{aligned}
p_{\mathrm{band}}^{2}
&=\int_{f_\mathrm{L}}^{f_\mathrm{U}} S_p(f)\,df \\[4pt]
&\approx \frac{2h}{45}\Bigl[
  7S_p\!\bigl(f^{(0)}\bigr)
 +32S_p\!\bigl(f^{(1)}\bigr)
 +12S_p\!\bigl(f^{(2)}\bigr)
 +32S_p\!\bigl(f^{(3)}\bigr)
 + 7S_p\!\bigl(f^{(4)}\bigr)
\Bigr] \\[4pt]
&= \frac{\Delta f}{90}\Bigl[
  7S_p\!\bigl(f^{(0)}\bigr)
 +32S_p\!\bigl(f^{(1)}\bigr)
 +12S_p\!\bigl(f^{(2)}\bigr)
 +32S_p\!\bigl(f^{(3)}\bigr)
 + 7S_p\!\bigl(f^{(4)}\bigr)
\Bigr].
\end{aligned}
\end{equation}
The corresponding one-third-octave band SL is then reported according to
\begin{equation}
\label{eq:band_sl}
L_{\mathrm{band}}
=
10\log_{10}\!\left(\frac{p_{\mathrm{band}}^{2}}{(1~\mu\mathrm{Pa})^{2}}\right),
\qquad
\left[\mathrm{dB\ re}\ 1~\mu\mathrm{Pa}^{2}\ @ 1~\mathrm{m}\right].
\end{equation}
A broadband SL over 20--2000~Hz is computed by summing the linear band energies across all one-third-octave bands whose band-edge frequencies lie within 20--2000~Hz, as given in
\begin{equation}
\label{eq:broadband_sl_sum}
p_{\mathrm{BB}}^{2}
=
\sum_{b\,:\,f_{b,\mathrm{L}}\ge 20,\ f_{b,\mathrm{U}}\le 2000} p_b^{2}
\end{equation}

\begin{equation}
\label{eq:broadband_sl}
L_{\mathrm{BB}}
=
10\log_{10}\!\left(\frac{p_{\mathrm{BB}}^{2}}{(1~\mu\mathrm{Pa})^{2}}\right)
\end{equation}

\subsubsection{Sound Exposure Level Estimation}
\label{sec:sel_estimation}

Sound exposure level (SEL) is a cumulative noise metric that represents the time-integrated acoustic energy received at a location over an observation window of duration \(T\). SEL is widely used to quantify exposure to underwater noise~\cite{hawkins2017sound, martin2019sound}. SEL is defined as
\begin{equation}
\label{eq:sel_def}
\mathrm{SEL}
=
10\log_{10}\!\left(
\frac{\displaystyle \int_{0}^{T} p^2(t)\,dt}{p_{\mathrm{ref}}^{2}\,t_{\mathrm{ref}}}
\right),
\qquad
\left[\mathrm{dB\ re}\ 1~\mu\mathrm{Pa}^2\,\mathrm{s}\right],
\end{equation}
where \(p(t)\) is the instantaneous acoustic pressure, \(p_{\mathrm{ref}} = 1~\mu\mathrm{Pa}\), and \(t_{\mathrm{ref}} = 1~\mathrm{s}\).

In \textit{ShipEcho}, SEL is estimated on a receiver grid from AIS-derived vessel activity over a user-defined observation window. Vessel activity is represented by AIS messages indexed by MMSI and timestamp. For SEL estimation, consecutive AIS reports for each vessel are used to construct time-weighted track segments, each of which is associated with a representative vessel position \(\mathbf{r}_s\), a duration \(\Delta t_s\), the vessel characteristics required by the selected SL model, and the frequency band of interest. In the current SEL implementation, the \textit{Combined} SL model is used and cannot be changed by the user. As defined in~\eqref{eq:combined_energetic_mean}, this model reports the energetic mean of the SL estimates from all available models for a given vessel in each indicator band. For each segment \(s\), a band SL, denoted \(L_{\mathrm{S},s}\), is first estimated using the \textit{Combined} model. This SL is then propagated from the vessel position to each receiver location \(\mathbf{x}\) using the Gaussian ray model, yielding a received level contribution \(L_{\mathrm{R},s}(\mathbf{x})\).

The received mean-square pressure contribution of segment \(s\) at receiver location \(\mathbf{x}\) is expressed in linear form
\begin{equation}
\label{eq:sel_seg_energy}
E_s(\mathbf{x})
=
p_{\mathrm{ref}}^{2}\,10^{L_{\mathrm{R},s}(\mathbf{x})/10}\,\Delta t_s.
\end{equation}
The cumulative sound exposure over the full observation window is then approximated by summing the contributions from all vessel track segments, as given below
\begin{equation}
\label{eq:sel_energy_sum}
E_{\mathrm{tot}}(\mathbf{x})
\approx
\sum_{s=1}^{S} E_s(\mathbf{x})
=
\sum_{s=1}^{S}
p_{\mathrm{ref}}^{2}\,10^{L_{\mathrm{R},s}(\mathbf{x})/10}\,\Delta t_s,
\end{equation}
where \(S\) is the total number of AIS-derived segments contributing within the selected time window.
Substituting~\eqref{eq:sel_energy_sum} into~\eqref{eq:sel_def} with \(t_{\mathrm{ref}}=1~\mathrm{s}\) yields the SEL estimate used in \textit{ShipEcho}, given by
\begin{equation}
\label{eq:sel_discrete}
\mathrm{SEL}(\mathbf{x})
=
10\log_{10}\!\left(
\sum_{s=1}^{S}
10^{L_{\mathrm{R},s}(\mathbf{x})/10}\,\Delta t_s
\right).
\end{equation}

\subsubsection{Propagation Modeling}
\label{subsec:propagation_model}

Estimating V-URN at a receiver location requires representation of acoustic propagation through the local marine environment. In \textit{ShipEcho}, propagation is modeled using a computationally efficient Gaussian ray-tracing (GRT) approach~\cite{porter1987_gaussian_beam_tracing,ameer2010localization,reilly2012sonar}, selected as a practical compromise between physical realism and computational cost for interactive large-area mapping.

The propagation model uses bathymetry and the sound-speed profile (SSP) as its main environmental inputs. Bathymetry is obtained from the General Bathymetric Chart of the Oceans (GEBCO) gridded dataset~\cite{gebco_gridded_bathymetry_data}, which defines the local seabed boundary and water depth for ray propagation and bottom interaction. The SSP is derived from seasonal-mean temperature and salinity fields from the World Ocean Atlas~\cite{NOAANCEI_WorldOceanAtlas} and converted to sound speed using Mackenzie’s empirical equation~\cite{Mackenzie1981_NineTermSoundSpeed}, given in
\begin{equation}
\begin{aligned}
c(T,S,D) &= 1448.96 + 4.591T - 5.304\times10^{-2}T^2 + 2.374\times10^{-4}T^3 \\
&\quad + 1.340(S-35) + 1.630\times10^{-2}D + 1.675\times10^{-7}D^2 \\
&\quad - 1.025\times10^{-2}T(S-35) - 7.139\times10^{-13}TD^3 ,
\end{aligned}
\label{eq:mackenzie_sound_speed}
\end{equation}
where $c$ is sound speed [m/s], $T$ is temperature [$^\circ$C], $S$ is salinity [ppt], and $D$ is depth [m].

For each vessel position, \textit{ShipEcho} launches rays over azimuth angles spanning $0^\circ$--$360^\circ$ and elevation angles spanning $-30^\circ$ to $+30^\circ$, using uniform angular steps of $\Delta \phi = 10^\circ$ in azimuth and $\Delta \theta = 5^\circ$ in elevation. These angular discretization settings are not prescribed by any formal standard and are instead model-specific parameters selected to balance spatial resolution and computational efficiency for interactive, large-area mapping. They provide full azimuthal coverage while concentrating elevation sampling on moderate-angle propagation paths that are expected to contribute most to received levels in the simplified propagation framework. The resulting ray fan contains $N_{\mathrm{az}} = 36$ azimuth directions and $N_{\mathrm{el}} = 13$ elevation angles, giving a total of $N_{\mathrm{ray}} = 468$ launched rays per vessel position. Each ray is propagated as a piecewise-linear path composed of straight segments between successive surface and seabed interactions, and the cumulative distance traveled from the source to an evaluation point along the ray path is denoted by $r_p$.

Received levels are evaluated on the map grid by computing, for each grid point, the minimum cross-ray distance $\rho$ to the ray path. To reduce computational cost, only grid points within a fixed neighborhood of the ray are considered, i.e., $\rho \leq 500$~m. Lateral spreading about the ray centerline is represented using a Gaussian beam weighting. The angular offset from the ray is defined as
\begin{equation}
\psi = \arctan\left(\frac{\rho}{r_p}\right),
\label{eq:ray_angle_offset}
\end{equation}
and the corresponding beam-weighting function is given by
\begin{equation}
w(\psi) = C \cos(\theta)\exp\left(-a\psi^2\right),
\label{eq:gaussian_beam_weight}
\end{equation}
where the parameter $a$ is defined by
\begin{equation}
a = -\frac{\ln(\beta)}{\Delta\theta^2},
\label{eq:gaussian_beam_a}
\end{equation}
with $\theta$ denoting the ray elevation angle, $\Delta\theta$ the elevation angular step expressed in radians, and $\beta$ a beam-shape parameter. In \textit{ShipEcho}, $\beta = 0.1$ is used as a practical compromise between spatial localization and overlap between neighboring rays. The normalization constant is taken as
\begin{equation}
C = \frac{1}{\sqrt{4\pi}},
\label{eq:gaussian_beam_norm}
\end{equation}
which provides isotropic amplitude normalization over $4\pi$ steradians.

Propagation is applied in transmission-loss form according to
\begin{equation}
\textrm{RL} = \textrm{SL} - \textrm{TL},
\label{eq:rl_sl_tl}
\end{equation}
where \textrm{RL} is the received level, \textrm{SL} is the source level, and \textrm{TL} is the transmission loss. Geometric spreading is represented by a spherical-spreading approximation given in
\begin{equation}
\textrm{TL}_{\mathrm{sp}} = 20\log_{10}(r_p),
\label{eq:spherical_spreading_tl}
\end{equation}
where $r_p$ is the cumulative source-to-receiver path length along the ray in meters.

Frequency-dependent absorption is modeled using Thorp’s empirical approximation~\cite{thorp1967ocean}. Defining $f$ as frequency in kHz, the absorption coefficient in dB/km is given by
\begin{equation}
\alpha(f) = \frac{0.11f^2}{1+f^2} + \frac{44f^2}{4100+f^2} + 2.75\times10^{-4}f^2 + 0.003,
\label{eq:thorp_absorption}
\end{equation}
and the corresponding absorption loss over path length $r_p$ is given by
\begin{equation}
\textrm{TL}_{\mathrm{abs}} = \frac{\alpha(f)}{1000}r_p,
\label{eq:absorption_tl}
\end{equation}
where the factor $1/1000$ converts dB/km to dB/m. Additional losses associated with sea-surface and seabed interactions are represented through per-interaction penalties applied along the ray path.

For each vessel $v$ and launched ray $r$, the propagation model yields an on-axis received level $\textrm{RL}_{v,r}(f)$ at frequency $f$ along the ray centerline, using the transmission-loss relationship in~\eqref{eq:rl_sl_tl}. To evaluate received levels on the map grid, energy is distributed laterally about each ray centerline using the Gaussian beam weighting defined by~\eqref{eq:gaussian_beam_weight}--\eqref{eq:gaussian_beam_norm}. For a grid point $\mathbf{x}$, the angular offset relative to the ray centerline is written as
\begin{equation}
\psi_{v,r}(\mathbf{x}) = \arctan\left(\frac{\rho(\mathbf{x})}{r_p}\right),
\label{eq:grid_ray_offset}
\end{equation}
where $\rho(\mathbf{x})$ is the minimum cross-ray distance from grid point $\mathbf{x}$ to the ray path. The accumulated mean-square pressure at $\mathbf{x}$ and frequency $f$ is then given by
\begin{equation}
p^2(\mathbf{x},f) = \sum_v \sum_r \left[w\!\left(\psi_{v,r}(\mathbf{x})\right)\right]^2 10^{\mathrm{RL}_{v,r}(f)/10},
\label{eq:mean_square_pressure_map}
\end{equation}
where the squared weighting term appears because $w(\psi)$ is an amplitude taper while accumulation is performed in the energy domain. The mapped received level is then expressed by
\begin{equation}
\mathrm{RL}_{\mathrm{map}}(\mathbf{x},f) = 10\log_{10}\left(\frac{p^2(\mathbf{x},f)}{(1~\mu\mathrm{Pa})^2}\right).
\label{eq:rl_map}
\end{equation}

The accumulation in~\eqref{eq:mean_square_pressure_map} is performed incoherently, consistent with the band-energy indicators reported by \textit{ShipEcho}. A coherent phase-resolved field is not constructed, since this would require wavelength-scale spatial sampling and substantially denser frequency sampling within each band, increasing both computational cost and sensitivity to small uncertainties in the propagation environment and vessel motion.

\subsubsection{AIS Integration}
\label{sec:ais_integration}

AIS is a very-high-frequency (VHF) radio-based maritime transceiver system that periodically broadcasts vessel information, including identifiers, vessel name and type, principal dimensions (length, beam, draft), and navigational parameters such as position, course, and speed. Under the International Maritime Organization (IMO) implementation of the Safety of Life at Sea (SOLAS) Convention, AIS carriage is mandatory for all ships of more than 300 gross tonnage on international voyages, cargo ships of more than 500 gross tonnage not engaged on international voyages, and all passenger ships, irrespective of size (commonly corresponding to Class~A AIS)\footnote{Smaller vessels (e.g., yachts and sailing vessels) may voluntarily carry AIS or may be required to do so under local regulations. These vessels typically use Class~B AIS units, which generally have lower transmission priority and longer nominal reporting intervals than Class~A systems~\cite{imo_ais}.}~\cite{imo_ais}.
Accordingly, AIS provides a practical basis for vessel position monitoring and for obtaining key operational parameters required for V-URN estimation.

\begin{figure}[!htbp]
\centering
\includegraphics[scale=0.15]{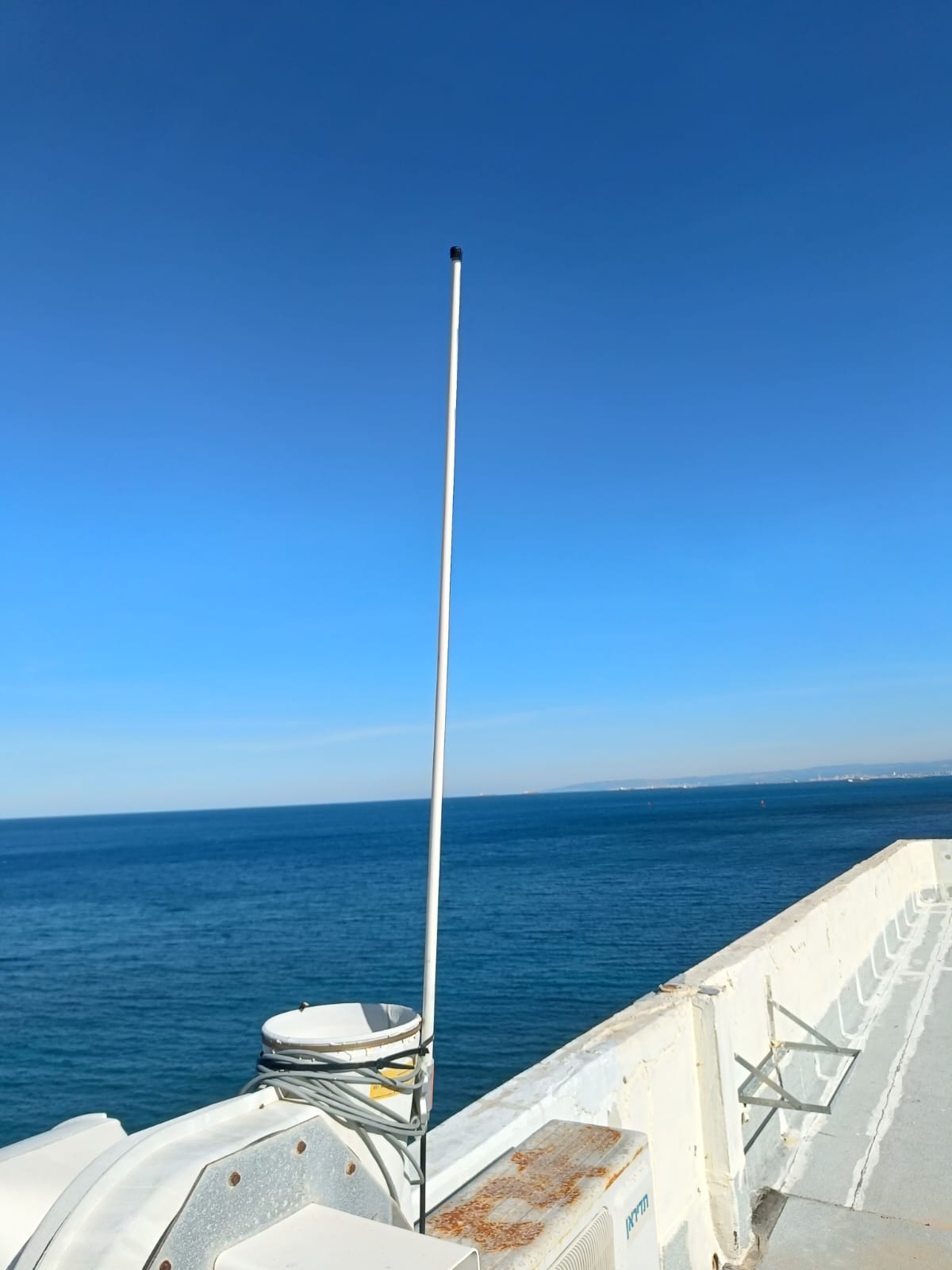}
\caption{AIS receiving station deployed near Haifa Port.}\label{AIS_Haifa}
\end{figure}

In practice, acquiring AIS data over large areas remains challenging. Terrestrial AIS reception is constrained by line-of-sight propagation, such that independently operated shore receivers typically provide only localized coverage; scaling to wider regions requires a dense, maintained receiver network. Alternatively, commercial AIS data service providers can offer broader coverage, both for online monitoring and for offline monitoring via past recordings; however, these services are often costly, which can limit their use in large-area operational applications. To address these constraints, \textit{ShipEcho} retrieves AIS data via an application programming interface (API) from AISHub, a community-based AIS data exchange that aggregates feeds from contributing stations~\cite{aishub}. Access to AISHub requires a contribution to expand AIS coverage. To that end, we deployed and operate two shore-based AIS receiving stations near the port entrances of Haifa and Ashdod in Israel. A picture of the AIS transceiver deployed near the port of Haifa is shown in Fig.~\ref{AIS_Haifa}.
Because AIS reporting intervals vary by equipment class and vessel dynamics, \textit{ShipEcho} provides a uniform update interval every $\Delta t=60$~s.

\subsection{Software Architecture}

\textit{ShipEcho} is implemented as a web-based client--server system in which the back-end provides standardized data services and the front-end supports interactive map-based analysis and visualization. The back-end is built with Node.js~\cite{nodejs_about} and Express.js~\cite{express_home}, and exposes a Representational State Transfer (REST) API for AIS data acquisition, processing, storage, export, and supporting geospatial and oceanographic computations. In operation, AIS messages are retrieved periodically from an external provider. To improve efficiency, a bounding box is first defined for each polygonal region and point-in-polygon filtering is then applied to retain only valid vessel positions, while excluded internal subregions such as inland waters or rivers are omitted. To reduce repeated external requests and improve responsiveness, the server maintains an in-memory cache of recent vessel states with a defined time-to-live (TTL), while persisting the same records in a MySQL database~\cite{oracle_what_is_mysql} to support historical and cumulative analyses. Bathymetric GeoTIFF rasters are loaded once at server startup and retained in memory to enable rapid depth lookup without repeated file access. Basic authentication is supported through registration and login, with passwords hashed using bcrypt~\cite{provos1999future} and registration protected by Google reCAPTCHA~\cite{google_recaptcha_docs}. Administrative endpoints provide user overview and account management. To manage the computational burden of SEL calculations, \textit{ShipEcho} applies tiered access limits, whereby guest users are restricted to single-day SEL calculations over relatively small domains, whereas registered users are permitted larger spatial domains and longer time windows.

The front-end is implemented as a single-page application in React~\cite{react_dev} and provides the primary user interface for exploring vessel traffic and derived V-URN metrics. Navigation between LVM, HM, and SELM is handled through React Router, enabling seamless transitions without full page reloads and allowing common interface components to be shared across modes. Spatial visualization is centered on an interactive MapLibre GL map~\cite{maplibre_gl_js_docs}, via React Map GL bindings, which renders vessel positions as dynamically scaled and rotated symbols and provides interactive popups containing AIS attributes and estimated noise metrics. To preserve performance under high vessel densities, the interface applies spatial filtering and clustering at lower zoom levels.

The system is designed for interactive use with frequent updates. At each update, the dominant computational cost is associated with propagation calculations and heatmap refresh over the selected domain, and this cost increases with the number of active vessels. Practical runtime is reduced by bounding the computational domain, caching recent vessel states, and reusing preloaded bathymetry. Application state and periodic updates are managed using React hooks, with memoization used to derive filtered vessel lists and heatmap inputs efficiently while minimizing unnecessary re-rendering. \textit{ShipEcho} also supports temporal analysis of historical AIS records through server-side queries combined with client-side playback controls, including date selection, timeline sliders, and configurable playback speed, so that only temporally valid vessels matching user-selected criteria are displayed at each time step. Users may additionally upload CSV datasets (see documentation in repository), which are parsed and normalized in the browser into the same internal structure used for API-delivered AIS records, thereby enabling a unified processing and visualization workflow regardless of data source.

\subsection{Validation}
\label{subsec:validation}

To provide an initial evaluation of the accuracy of \textit{ShipEcho} V-URN maps, modeled daily SEL estimates were compared with calibrated passive acoustic measurements collected near the main shipping route to the Port of Haifa, Israel, at $32^\circ53'58.8''$~N, $34^\circ54'17.8''$~E. The measurements were acquired using an RTSYS EA-SDA14 recorder coupled with a COLMAR GP1190 omnidirectional hydrophone with a sensitivity of $-178$~dB re~1~V/$\mu$Pa. The hydrophone was deployed at a water depth of approximately 66~m and positioned about 4~m above the seabed.

Because \textit{ShipEcho} estimates AIS-tracked vessel contributions only, the calibrated recordings were expected to include additional sound sources not explicitly represented by the model, including AIS-exempt or AIS-inactive small craft, biological sound, and wind- and wave-generated ambient noise. The analysis focused on two representative days, 6~August~2025 and 7~August~2025. Daily SEL values derived from \textit{ShipEcho} and from the calibrated recordings were evaluated in the standard indicator bands, namely the one-third-octave bands centered at 63 and 125~Hz, together with the broadband 20--2000~Hz metric. The comparison was performed using the nearest \textit{ShipEcho} grid cell to the recorder location. In this case, the recorder position was located near the center of the selected grid cell, reducing point-to-cell mismatch effects. The comparison results are summarized in Table~\ref{tab:validation-sel-comparison}.

\begin{table}[!htbp]
\centering
\caption{Comparison of measured daily SEL and \textit{ShipEcho} estimates for the Haifa validation case. Differences are calculated as \textit{ShipEcho} minus measured SEL.}
\label{tab:validation-sel-comparison}
\small
\setlength{\tabcolsep}{4pt}
\renewcommand{\arraystretch}{1.15}
\begin{tabular}{llccc}
\hline
\textbf{Date} & \textbf{Source} & \textbf{63 Hz} & \textbf{125 Hz} & \textbf{20--2000 Hz} \\
\hline
06.08.2025 & Measured & 152.62 & 153.09 & 165.58 \\
06.08.2025 & \textit{ShipEcho} & 152.68 & 144.83 & 159.86 \\
06.08.2025 & Difference & 0.06 & -8.26 & -5.72 \\
\hline
07.08.2025 & Measured & 157.62 & 158.05 & 170.57 \\
07.08.2025 & \textit{ShipEcho} & 161.69 & 154.29 & 168.43 \\
07.08.2025 & Difference & 4.07 & -3.76 & -2.14 \\
\hline
\end{tabular}
\end{table}

The Haifa validation case showed good first-order agreement between \textit{ShipEcho} and the calibrated daily SEL measurements, particularly considering that the two datasets do not represent exactly the same acoustic quantity. On 6~August~2025, \textit{ShipEcho} reproduced the measured 63~Hz daily SEL very closely, with a difference of only 0.06~dB. Larger underestimates were observed at 125~Hz and for the 20--2000~Hz broadband metric, with differences of -8.26 and -5.72~dB, respectively. On 7~August~2025, \textit{ShipEcho} overestimated the 63~Hz daily SEL by 4.07~dB, while underestimating the 125~Hz and broadband metrics by 3.76 and 2.14~dB, respectively.

Overall, the agreement was strongest at 63~Hz, where the modeled and measured daily SEL values were within 0.1~dB on 6~August and within approximately 4~dB on 7~August. The remaining differences are consistent with the expected limitations of an AIS-based vessel-noise model. \textit{ShipEcho} estimates the contribution from AIS-tracked traffic, whereas the hydrophone recorded the complete local soundscape, including ambient sea noise, weather-driven wave noise, biological sound, AIS-exempt or AIS-inactive small craft, and other non-modeled sources. These additional components can increase the measured SEL relative to the modeled value, especially in the broadband metric, and may also influence the 63 and 125~Hz indicator bands depending on local traffic composition and environmental conditions.

The differences reported in Table~\ref{tab:validation-sel-comparison} therefore reflect the combined influence of SL uncertainty, incomplete AIS representation of local traffic, and the inherent difference between an AIS-based modeled vessel-noise estimate and a calibrated measurement of the full underwater soundscape. Despite these limitations, the Haifa comparison indicates that \textit{ShipEcho} can reproduce daily SEL levels to within a few decibels when the model grid cell and hydrophone position are well aligned.

\section{Potential of \textit{ShipEcho} as a marine management tool}
\textit{ShipEcho} is designed as a management-level decision-support tool that translates vessel activity into interpretable, map-based V-URN estimates for environmental assessment and spatial planning. By combining AIS-derived vessel positions and operating parameters with environmental data and established acoustic SL and propagation models, \textit{ShipEcho} provides an accessible GIS interface that enables users to explore how predicted noise levels vary across space and time. This enables rapid screening of potential high-exposure areas (e.g., near ports and shipping lanes) and supports consistent reporting using widely adopted monitoring indicators. In particular, \textit{ShipEcho} reports levels in the low-frequency continuous-sound indicator bands centered at 63~Hz and 125~Hz (one-third octave), which are commonly used for assessing shipping-dominated ambient noise under the Marine Strategy Framework Directive (MSFD) Descriptor~11 framework~\cite{ec2017_848_msfd_descriptor11}. In addition, it provides a 20--2000~Hz broadband metric that summarizes a large fraction of biologically relevant soundscape energy and is frequently used in ecological soundscape analyses~\cite{vieira2021meagre}.
To align with practical management workflows, \textit{ShipEcho} emphasizes transparency and exploratory analysis over static reporting. Users can compare alternative SL models, visualize the resulting differences in predicted V-URN distributions, and interpret outcomes in relation to spatial management objectives. Importantly, \textit{ShipEcho} overlays MPA boundaries within the map display, enabling direct examination of predicted noise exposure within and around legally designated protected zones. Through its real-time, historical, and SEL views, \textit{ShipEcho} supports both operational situational awareness and retrospective assessment, providing a flexible platform for managers and stakeholders to contextualize vessel-noise exposure and evaluate mitigation or planning scenarios.
In the following section, we provide an example use of \textit{ShipEcho} - estimation of sound exposure levels in marine protected areas, and the effect of speed reduction on sound exposure levels.

\subsection{SEL Assessment in an MPA and Evaluation of a Slowdown Scenario}
\label{sec:sel_mpa_slowdown}

To illustrate \textit{ShipEcho} as a management-scale decision-support tool, we estimate SELs within the Jabuka Basin Reef (Grebeni u Jabu\v{c}koj kotlini) MPA, Croatia, between 1--7~July~2025. We then evaluate the potential effect of a speed-reduction (slowdown) region, similar in concept to slowdown programs implemented in Admiralty Inlet (Washington, USA)~\cite{quietsound_admiralty_slowdown_2026} and the Haro Strait region (Canada)~\cite{echo_program_2026_slowdowns}.

\begin{figure}[!htbp]
\centering
\includegraphics[width=0.7\textwidth]{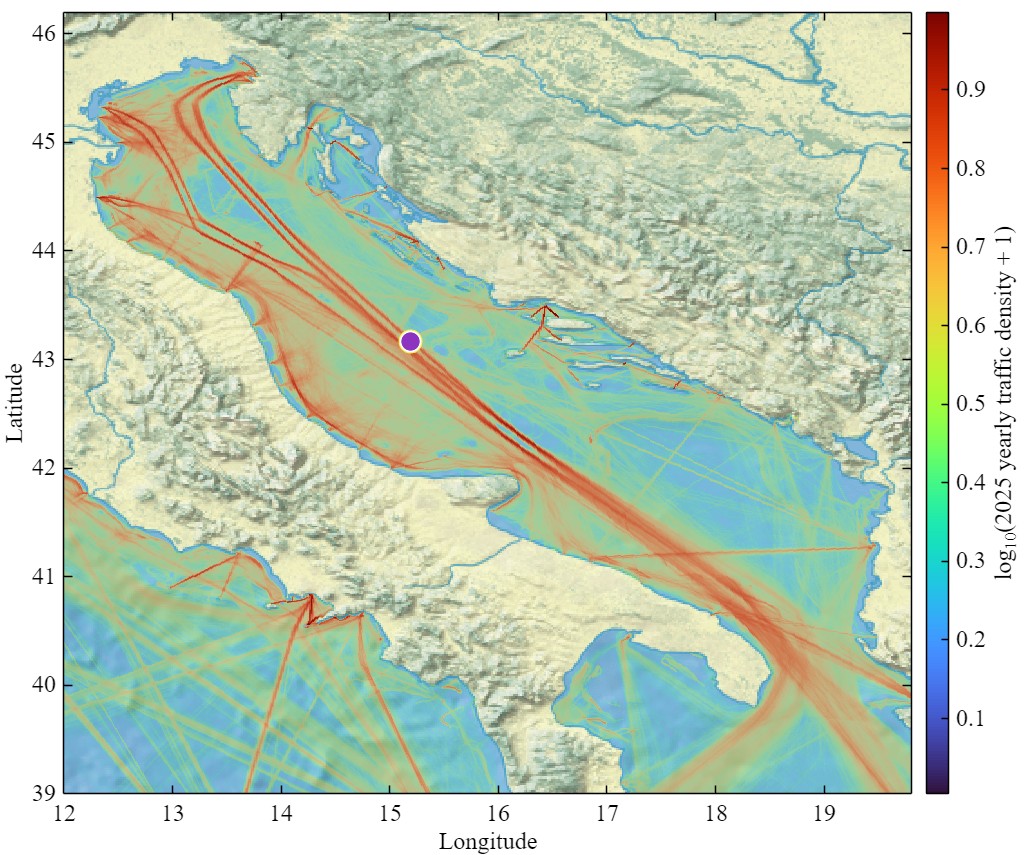}
\caption{Shipping density in the Adriatic Sea, retrieved from the European Marine Observation and Data Network~\cite{EMODnetVesselDensity}. The Jabuka Basin Reef MPA is indicated by a purple circle.}\label{fig:shipping_density_adriatic}
\end{figure}

The Jabuka Basin Reef MPA lies approximately 0.5~km north of Jabuka Island and is designated under the EU Habitats Directive. It covers \(\sim 11~\mathrm{km}^2\) and is characterized by distinct bathymetric features~\cite{EEA_EUNIS_HR3000477}. The site provides a useful test case due to its close proximity to central Adriatic shipping lanes connecting major regional ports (e.g., Trieste and Rijeka) with the Mediterranean Sea~\cite{luvsic2017analysis}, the location of the MPA and shipping density is provided in Fig.~\ref{fig:shipping_density_adriatic}. 

\begin{figure}[!htbp]
\centering
\includegraphics[width=0.9\textwidth]{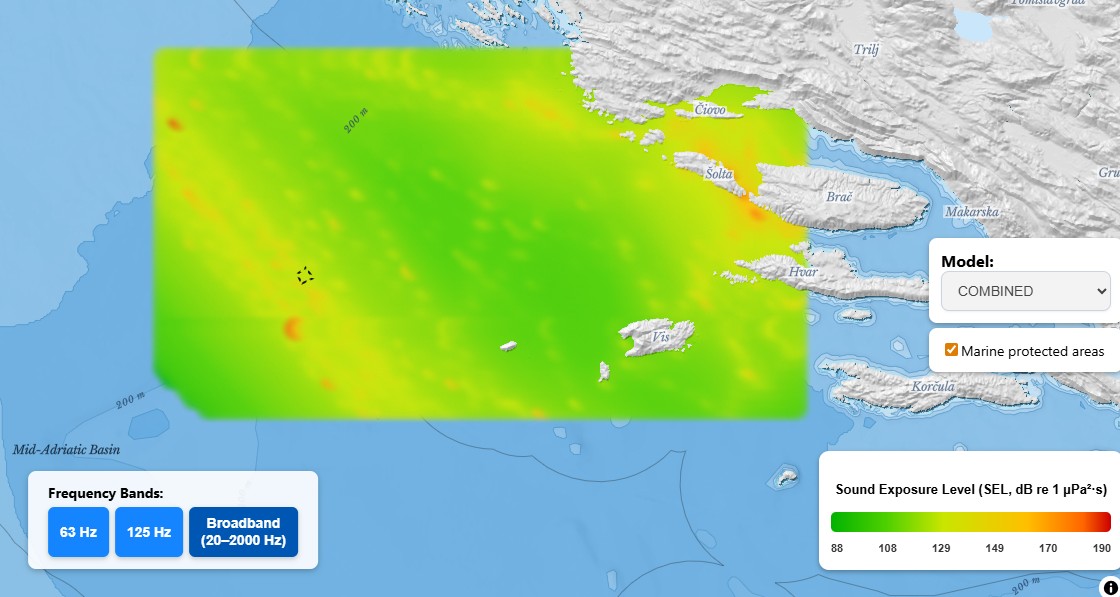}
\caption{SEL map generated by \textit{ShipEcho} for the area surrounding the Jabuka Basin Reef MPA for 1--7~July~2025 in the 20--2000~Hz band. The MPA boundary is indicated by a black dashed polygon. Higher SEL values are concentrated along nearby shipping routes.}
\label{fig:SEL_MPA_WEEK}
\end{figure}

Using SELM in \textit{ShipEcho}, SEL maps were generated for the area surrounding the MPA. To contextualize short-term variability in exposure within the MPA, AIS data were used to summarize large-vessel traffic within 20~km of the MPA and to describe the associated speed statistics. This radius was selected as a pragmatic local-traffic attribution scale that captures vessels most likely to dominate day-to-day SEL variability within the MPA; similar assessments have found that commercial vessels within approximately 20~km are typically the primary contributors to short-term SEL metrics~\cite{merchant2012_assessing,haver2021_largevessel}. Figure~\ref{fig:SEL_MPA_WEEK} shows the \textit{ShipEcho} SEL calculation for the full evaluation period, with the Jabuka Basin Reef MPA denoted by a dashed black polygon. Higher SEL values are concentrated along nearby shipping routes.

\begin{figure}[!htbp]
\centering
\includegraphics[width=1\linewidth]{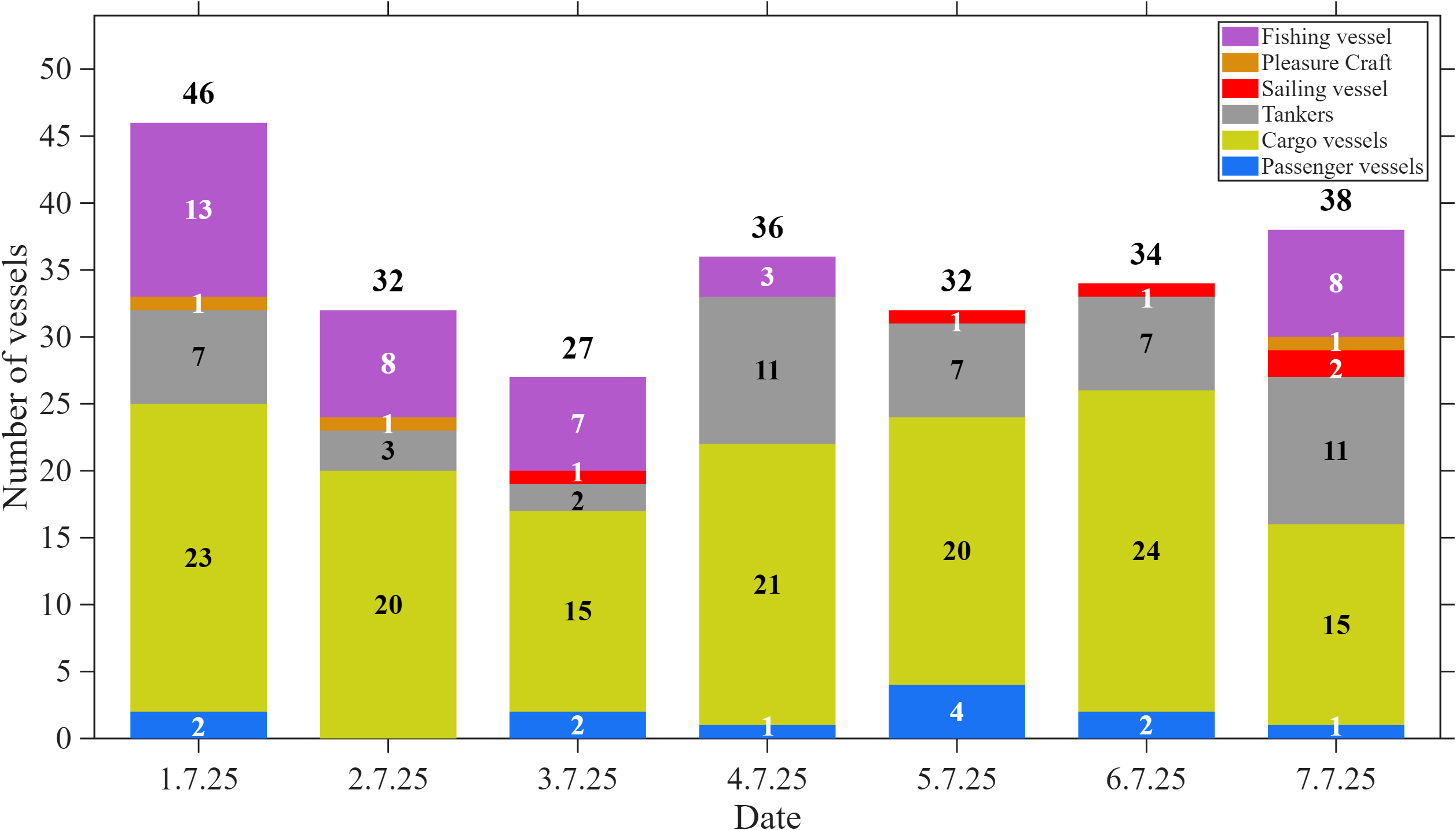}
\caption{Daily vessel transits within 20~km of the Jabuka Basin Reef MPA, grouped by vessel type, for 1--7~July~2025.}
\label{fig:vessel_count_perday}
\end{figure}

\begin{figure}[!htbp]
\centering
\includegraphics[width=1\linewidth]{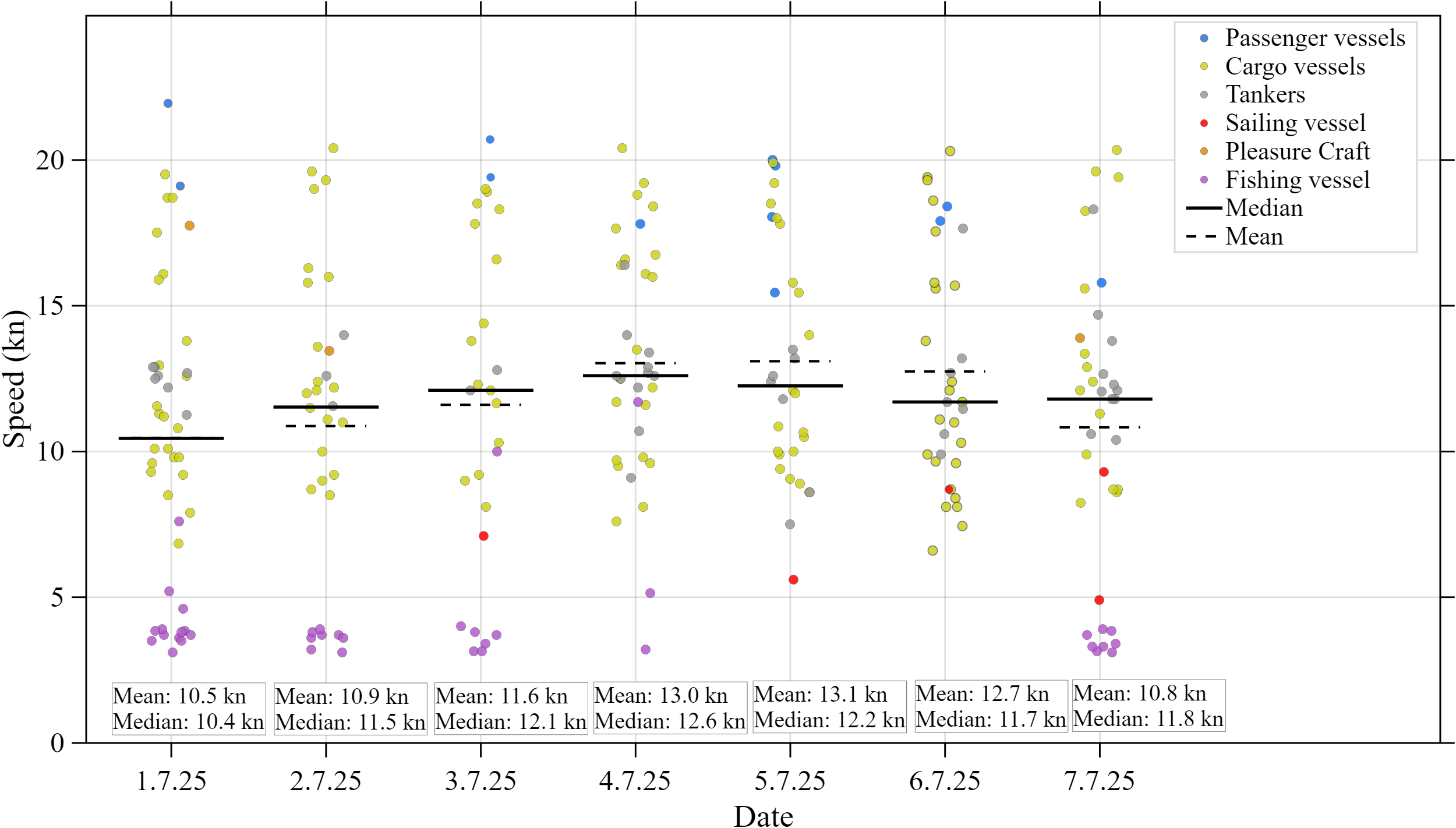}
\caption{Daily vessel speeds within 20~km of the Jabuka Basin Reef MPA, shown by vessel type, with daily mean and median speeds for 1--7~July~2025.}
\label{fig:mpa_speed_perday}
\end{figure}

Figure~\ref{fig:vessel_count_perday} reports daily passage counts for vessels within 20~km of the Jabuka Basin Reef MPA according to vessel type, while Fig.~\ref{fig:mpa_speed_perday} shows daily vessel-speed distributions together with the daily mean and median speeds.

As shown in Fig.~\ref{fig:vessel_count_perday}, vessel traffic within 20~km of the MPA was dominated by cargo vessels and tankers, with an average of 26 vessels per day transiting the main Adriatic shipping routes between Trieste, Rijeka, and the Mediterranean. Additional traffic consisted mainly of limited fishing-vessel activity and sporadic passenger-vessel transits.

As shown in Fig.~\ref{fig:mpa_speed_perday}, cargo vessels and tankers generally transited the area at speeds between 7 and 21~kn, while the few passenger vessels passed through the area at speeds between 16 and 23~kn. Fishing vessels showed the slowest transit speeds, typically between 3 and 4~kn. Accordingly, cargo-vessel and tanker transits are estimated to be the main drivers of V-URN within the MPA area. 

\begin{figure}[!htbp]
\centering
\includegraphics[width=0.9\textwidth]{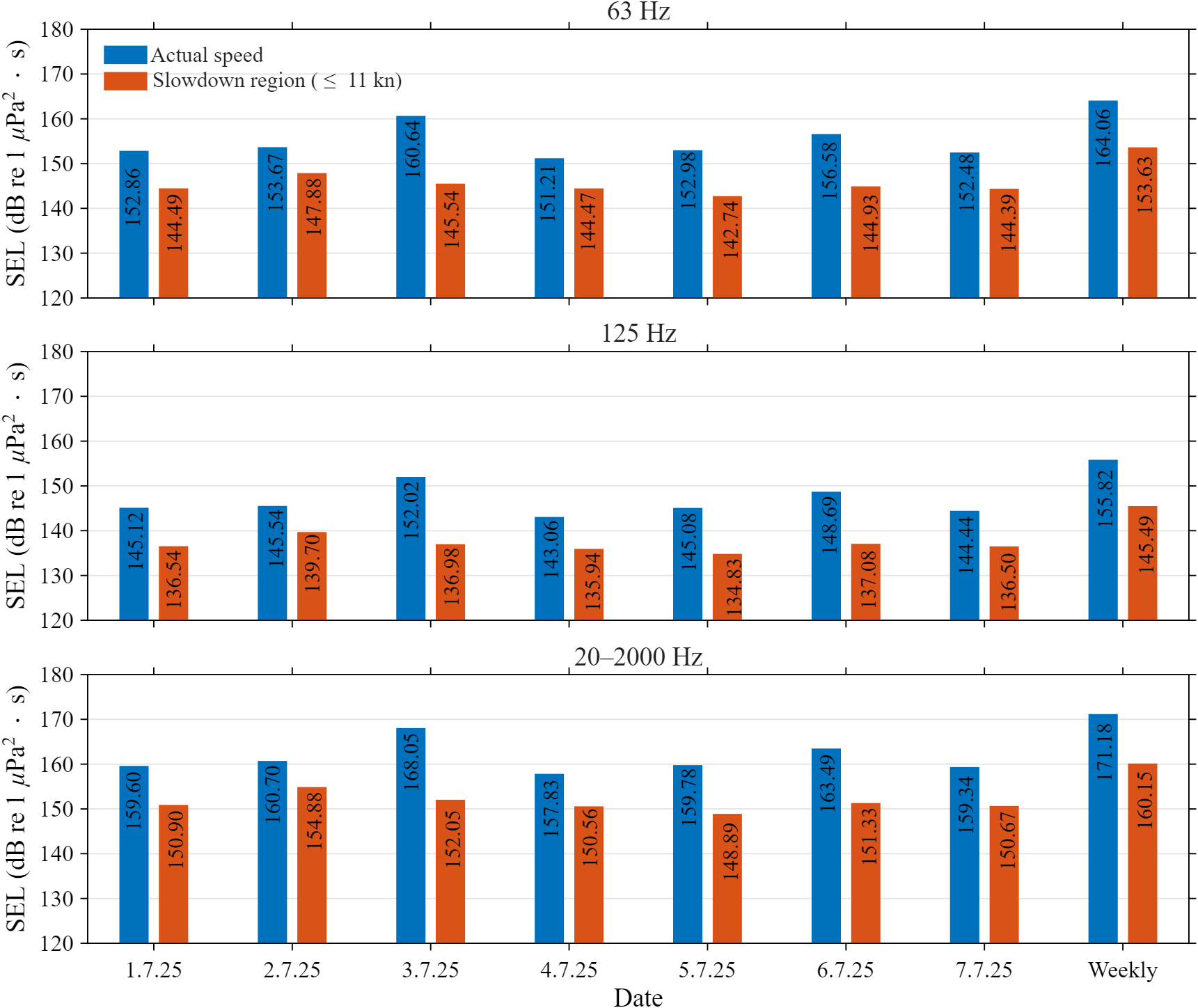}
\caption{Energetic-mean daily SEL within the Jabuka Basin Reef MPA for 1--7~July~2025 under observed vessel speeds (blue) and a simulated slowdown region with a speed cap of 11~kn within 20~km of the MPA (orange), shown for the 63~Hz and 125~Hz one-third-octave bands and the 20--2000~Hz broadband metric.}
\label{fig:sel_compare}
\end{figure}

To test the potential effect of speed reductions on daily SEL, daily energetic-mean SEL values within the MPA were computed for two scenarios: observed vessel speeds and a simulated slowdown region\footnote{To simulate the slowdown region, vessel speeds within 20~km of the MPA were capped at 11~kn; vessels already traveling below 11~kn were unchanged.}. Fig.~\ref{fig:sel_compare} compares the resulting daily SEL estimates for each indicator band.

Fig.~\ref{fig:sel_compare} shows that applying an 11~kn speed cap within 20~km of the MPA produces consistent reductions in energetic-mean SEL over the 1--7~July~2025 period. At the weekly scale, the energetic-mean SEL decreases from 164.06 to 153.63~dB at 63~Hz (\(\Delta = 10.43\)~dB), from 155.82 to 145.49~dB at 125~Hz (\(\Delta = 10.33\)~dB), and from 171.18 to 160.15~dB for the 20--2000~Hz broadband metric (\(\Delta = 11.03\)~dB). On a day-to-day basis, the reduction ranges from 5.79 to 15.10~dB at 63~Hz, from 5.84 to 15.04~dB at 125~Hz, and from 5.82 to 16.00~dB for the 20--2000~Hz broadband metric.

The largest daily reduction was observed on 3~July~2025. Under the simulated speed-cap scenario, SEL decreased from 160.64 to 145.54~dB at 63~Hz, from 152.02 to 136.98~dB at 125~Hz, and from 168.05 to 152.05~dB for the 20--2000~Hz broadband metric. As shown in Fig.~\ref{fig:vessel_count_perday}, this date exhibited the strongest SEL decrease despite having the lowest number of vessel transits within 20~km of the MPA during the week. Moreover, the daily vessel speeds were not substantially different from those on the other days, as shown in Fig.~\ref{fig:mpa_speed_perday}. The main distinguishing factor was instead the close proximity of several vessel passages to the MPA. On this date, two relatively high-speed cargo vessels (17.9 and 18.5~kn) passed inside the MPA, two additional cargo vessels passed within less than 0.7~km of the MPA boundary, and two passenger vessels passed within approximately 1--2~km. These close transits provide a plausible explanation for both the elevated SEL values before mitigation and the pronounced reduction after the speed cap was applied.
Overall, these results indicate that a localized vessel-speed reduction could substantially reduce cumulative acoustic exposure within the MPA over both daily and weekly timescales, and further illustrate how \textit{ShipEcho} can be used to evaluate prospective management measures.

\section{Discussion}
\label{sec:discussion}

\textit{ShipEcho} was developed as a management and decision-support tool for translating vessel activity into interpretable, map-based V-URN metrics for environmental assessment and marine spatial planning. By integrating AIS-derived vessel activity with environmental inputs and established SL and propagation models, the framework enables rapid screening of areas and periods where elevated acoustic exposure is likely to occur. Its main contribution is therefore not the implementation of a high-fidelity acoustic solver, but the development of an operational and scalable GIS-based framework for exploring spatio-temporal V-URN patterns at decision-maker scales. This capability is particularly relevant for routine assessments requiring frequent updates across large numbers of vessels, such as the case study presented in Section~\ref{sec:sel_mpa_slowdown}, where daily SEL within a designated MPA is evaluated under baseline and speed-management scenarios.

A further contribution is the ability to support exploratory, scenario-driven analysis rather than static reporting. \textit{ShipEcho} allows users to switch between alternative vessel SL models and directly assess how these assumptions influence mapped exposure fields and derived indicators. This is important in policy and management contexts, where uncertainty in input assumptions must be made explicit. By linking indicator computation to interactive map-based visualization, the framework supports identification of spatial hotspots, interpretation of temporal variability, and communication of scenario-based differences in a transparent and reproducible manner.

These strengths should be considered alongside several limitations that constrain the interpretation of \textit{ShipEcho} outputs. First, the accuracy of AIS-derived noise maps depends on the completeness and quality of vessel-activity data. AIS coverage varies geographically and is influenced by reception conditions, transmitter class, reporting rates, and carriage requirements, which may result in underrepresentation of smaller vessels and introduce spatial and class-dependent biases. Second, SL estimates are derived from parametric models whose applicability depends on the vessel categories and operating conditions represented in their underlying datasets. Consequently, absolute levels, and in some cases relative differences between scenarios, should be interpreted as indicative unless supported by local validation. Third, the propagation approach is optimized for computational efficiency to support interactive mapping across large spatial domains. While this enables rapid scenario evaluation, it does not fully resolve site-specific processes in complex coastal environments, including range-dependent boundary losses, detailed seabed geoacoustics, strong sea-state dependence, and coherent multipath structure. As a result, compared with high-resolution site-specific acoustic models, \textit{ShipEcho} trades physical completeness for speed, repeatability, and interpretability.

Accordingly, \textit{ShipEcho} is best suited for screening, prioritization, and comparative scenario analysis, such as baseline versus mitigation measures, as well as for transparent communication of assumptions and uncertainties. It is not intended to replace measurement-based assessments or high-fidelity acoustic modeling where regulatory decisions require tighter uncertainty bounds. The validation against calibrated measurements in Haifa supports this intended role. The comparison showed that \textit{ShipEcho} reproduced daily SEL levels within a few decibels for the Haifa case, particularly where the hydrophone position and model grid cell were well aligned. However, the remaining differences also highlight the main limitations of the framework. \textit{ShipEcho} estimates the AIS-tracked vessel-noise contribution, whereas hydrophones measure the complete underwater soundscape, including ambient sea noise, biological sound, AIS-exempt or AIS-inactive vessels, and other non-modeled sources. Therefore, validation results should be interpreted as an assessment of how well the modeled AIS-based vessel-noise field agrees with measured sound exposure under favorable spatial conditions, rather than as a full reconstruction of the measured soundscape.

Where local acoustic measurements are available, \textit{ShipEcho} is intended to complement monitoring by supporting calibration and validation, and by enabling iterative refinement of model assumptions in a reproducible framework. Although the system is global in design, operational constraints currently limit persistent storage and computationally intensive products, such as SEL aggregation, to predefined coastal regions, with inland waters excluded. This reflects a practical trade-off between computational cost and interactive performance, and can be addressed through incremental expansion or increased back-end capacity.

Future development can further enhance the utility of \textit{ShipEcho} for marine environmental management. One priority is the incorporation of an explicit ambient-noise component driven by metocean conditions, allowing vessel noise to be interpreted within the broader soundscape and enabling assessments based on exceedance or risk. A second priority is the integration of complementary vessel-detection systems, such as coastal radar or satellite-based observations, to improve representation of non-AIS traffic and reduce bias in coastal regions. Finally, the development of automated reporting tools capable of generating standardized temporal summaries for user-defined areas, including MPAs, port approaches, and shipping corridors, would support routine monitoring, stakeholder communication, and adaptive management.

Overall, \textit{ShipEcho} provides a practical bridge between vessel-activity data, acoustic modeling, and marine management needs. Its value lies in enabling rapid, transparent, and repeatable assessment of relative V-URN patterns and mitigation scenarios, while clearly identifying where local measurements or higher-fidelity modeling are required for more detailed site-specific evaluation.

\section*{CRediT authorship contribution statement}

\textbf{Mark Shipton:} Conceptualization, Methodology, Formal analysis, Writing -- original draft.  

\textbf{Valentino Denona:} Implementation, Software, Writing -- original draft.  

\textbf{Đula Nađ:} Project administration, Writing -- review \& editing, Resources.  

\textbf{Roee Diamant:} Supervision, Conceptualization, Project administration, Writing -- review \& editing, Resources.

\section*{Declaration of competing interest}
The authors declare that they have no known competing financial interests or personal relationships that could have appeared to influence the work reported in this paper.

\clearpage
\newpage

\appendix

\section{User Interface}
\label{app1}

\subsection{Live vessel mode (LVM)}
\label{LVM}
\begin{figure}[!htbp]
\centering
\includegraphics[width=1\textwidth]{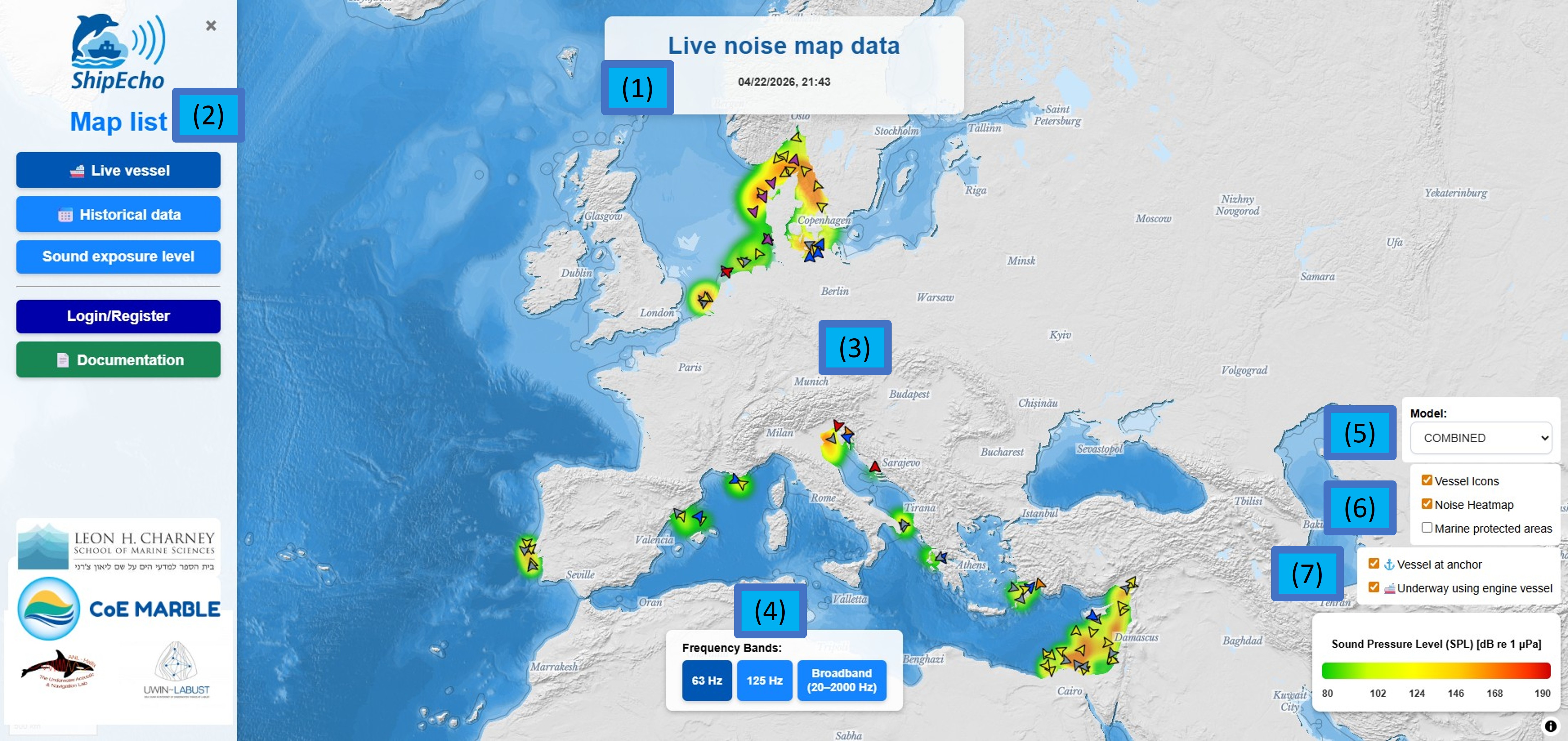}
\caption{An example of LVM over the area of Europe.}\label{fig:LVM}
\end{figure}

LVM is the default landing page. An example screenshot is shown in Fig.~\ref{fig:LVM}.

\begin{figure}[!htbp]
\centering
\includegraphics[width=1\textwidth]{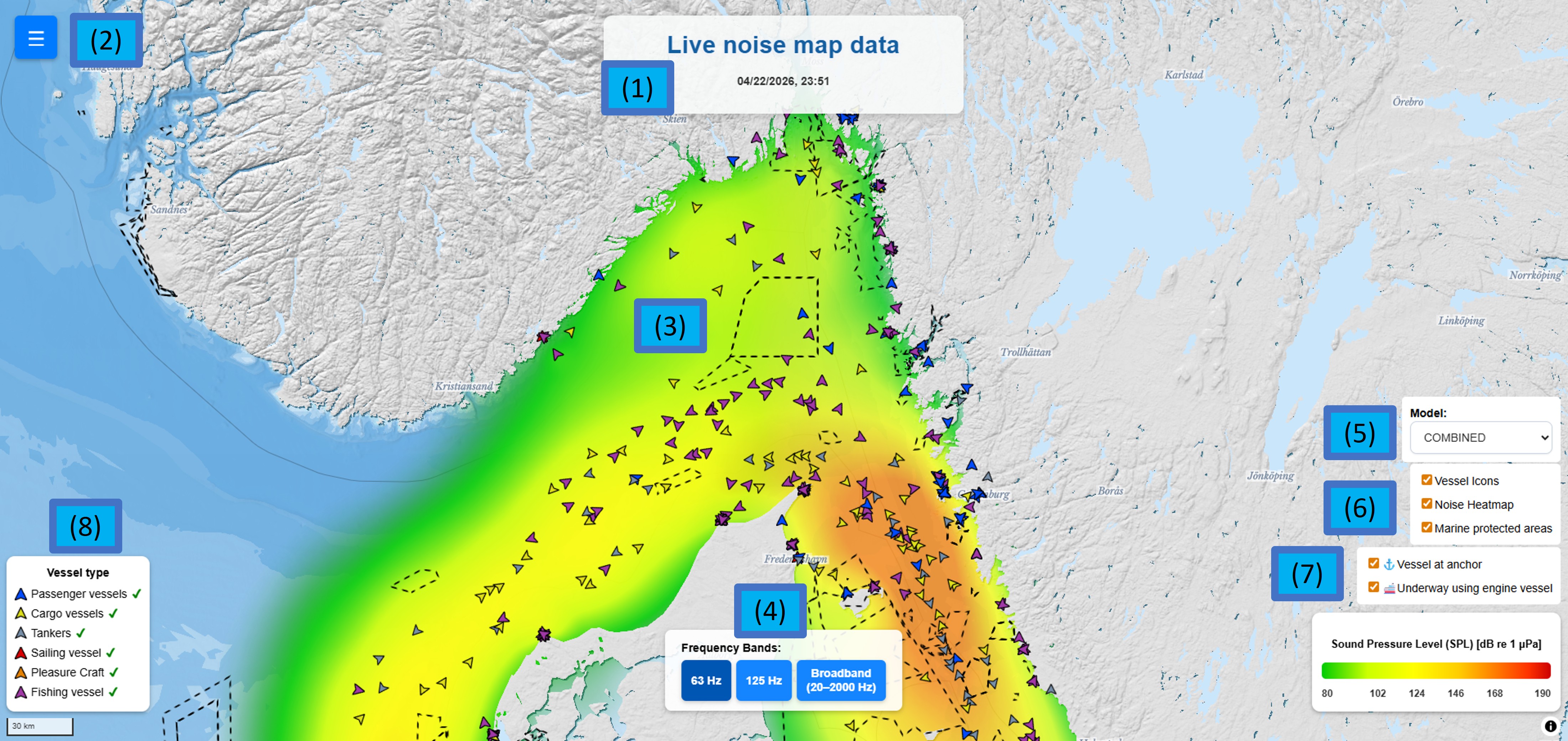}
\caption{An example of LVM with the MPA overlay enabled.}\label{fig:LVM_zoom}
\end{figure}

The header (1) reports the active display mode and the timestamp of the most recent AIS update. The left sidebar (2) provides navigation between display modes, user login/registration, and access to documentation; it can be minimized to increase the map view. The map and estimated URN field (3) form the main display and support pan and zoom interaction. For wide-area views (zoom levels 1--8), the number of displayed vessels is reduced to limit computation time; at higher zoom levels ($>8$), additional vessel positions are shown (Fig.~\ref{fig:LVM_zoom}). At the bottom (4), users select the frequency band (63~Hz, 125~Hz, or 20--2000~Hz). The display updates automatically after selection. In LVM, V-URN is reported as sound pressure level (SPL) \(\mathrm{dB\ re}\ 1~\mu\mathrm{Pa}\) and is color-coded according to the scale shown in the lower-right corner.

On the mid-right (5), users select the vessel SL model, including RANDI~3.1, JOMOPANS-ECHO, LBDS, AQUO, SRV, and Combined.
The display updates automatically. Below this (6), map layers can be enabled or disabled. Deselecting \textit{Show vessel icons} removes vessel symbols, deselecting \textit{Show noise heatmap} removes the propagated noise field, and enabling \textit{Marine protected areas} overlays legally designated MPAs from the WDPA dataset, allowing URN levels to be inspected within protected regions. The check boxes (7) filter vessels by AIS navigational status. \textit{Vessel at anchor} corresponds to AIS status 1, while \textit{Underway using engine} corresponds to status '0'\footnote{For a list of AIS status codes see~\cite{ITU_M1371_5_2014}. Vessels with status codes other than 0 or 1 are not displayed in \textit{ShipEcho}.}.

An LVM view of the area of Skagerrak (between Denmark and Norway), with the sidebar minimized and the MPA overlay enabled, is shown in Fig.~\ref{fig:LVM_zoom}.

MPAs are shown as polygons with black dashed outlines. With the sidebar minimized (2), the vessel legend (8) is visible and summarizes vessel categories and icons. A green checkmark indicates vessel types supported by the selected SL model (Table~\ref{tab:model-ais-types}). The sidebar can be restored using the icon (2).

\subsubsection{Historical Mode (HM)}
\label{HM}
\begin{figure}[!htbp]
\centering
\includegraphics[width=1\textwidth]{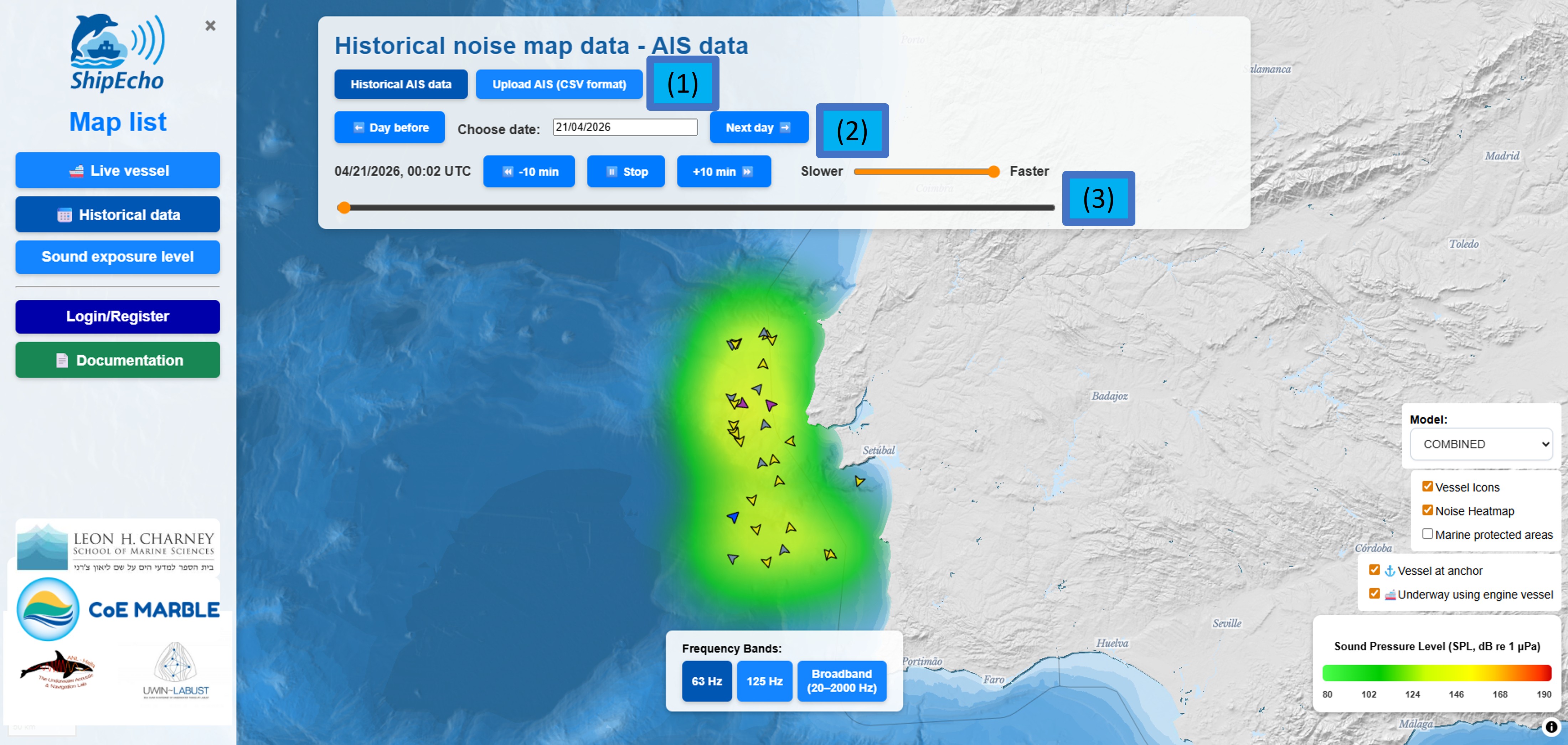}
\caption{An example of HM in western Portugal.}\label{fig:HM}
\end{figure}

HM enables playback of archived AIS data for specified dates and times. An example display is shown in Fig.~\ref{fig:HM}.

In the header (1), users select whether to replay AIS data stored on \textit{ShipEcho} or to upload a user-provided AIS dataset\footnote{The required file format and examples are provided in the repository.}. Users then select a start date (only valid dates are available). Pressing \textit{Start} begins playback. As in LVM, the displayed V-URN depends on the selected frequency band and SL prediction model. Users can step by $\pm$10~min or jump to the next day using the relevant buttons, and can adjust the displayed time using the timeline slider (2). Playback speed is controlled using an additional slider (3).

\subsubsection{Sound Exposure Level Mode (SELM)}
\label{SELM}
\begin{figure}[!htbp]
\centering
\includegraphics[width=1\textwidth]{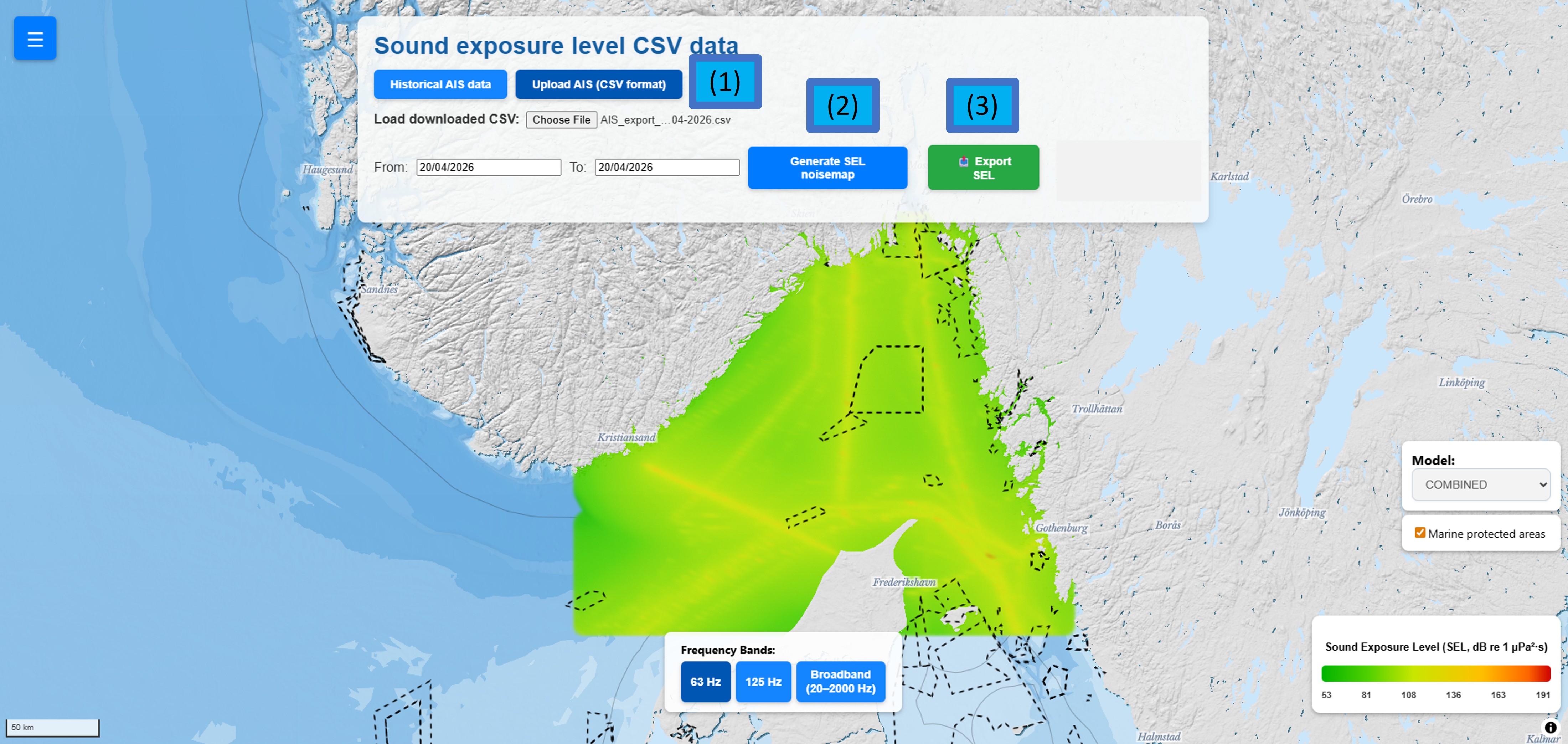}
\caption{Example of an SELM view with MPA overlay for the Skagerak area in the 63~Hz one-third-octave band for the 20-~April~2026.}\label{fig:SEL}
\end{figure}

SELM provides sound exposure level (SEL) estimates over user-defined areas and time intervals. An example display is shown in Fig.~\ref{fig:SEL}.

As in HM, users select whether to use AIS data stored on \textit{ShipEcho} or to upload AIS data (1). To generate an SEL map, users specify a date range and select an area of interest\footnote{To manage computation resources, guest users (unregistered) are limited to SEL calculations over a single day and an area of $2^\circ \times 2^\circ$ ($\approx 4.96\times10^{4}~\mathrm{km}^2$), while registered users are limited to time spans $<30$ days and an area of $10^\circ \times 10^\circ$ ($\approx 1.24\times10^{6}~\mathrm{km}^2$).}. Pressing \textit{Generate SEL noise map} (2) computes SEL over the selected region. Results can be downloaded using \textit{Export SEL} (3)\footnote{Downloaded outputs are provided as a CSV file containing SEL values for the selected grid cells, indexed by latitude and longitude. An example export file is included in the repository.}

\section{Detailed Description of Vessel SL Models}
\label{app2}

\subsection{The RANDI 3.1 Model}
The RANDI 3.1 model~\cite{breeding1996randi} utilizes a similar formulation to that of Ross and Alvarez~\cite{RossAlvarez1964RadiatedNoise}. The model is formalized by

\begin{equation}
\label{EQ.1}
\mathrm{SL}(f) = 190.5 
+ 50 \log_{10}\!\left(\frac{v}{10}\right) 
+ 20 \log_{10}\!\left(\frac{l}{150}\right) 
- 20 \log_{10}(f)
\end{equation}
where $\mathrm{SL}(f)$ is the estimated SL at a specific frequency $f$, $v$ is the speed of the vessel in knots, $l$ is the length of the vessel in meters. The equation defines reference parameters, which act as a baseline representing the characteristics of an 'average' vessel\footnote{The characteristics of an 'average vessel', i.e., length of 150 m at 10 kn, were based on vessel characteristics at the time (1950s-1960s).} where an increase in speed ($v$) or length ($l$) also increases the second and third terms of the equation, respectively, and vice versa.

The RANDI 3.1 model uses a similar formulation, albeit with several differences - different reference length and speed values are used (300 ft $\approx$ 91.4 m instead of 150 m and 12 knots), length and frequency dependent correction factors are used, and a frequency dependent baseline spectrum is used instead of the 190.5 value constant, such that
\begin{equation}
\label{Eq:RANDI}
\begin{aligned}
\mathrm{SL}_{1\,\mathrm{Hz}}(f) ={}& L_{\mathrm{S},0}(f)
+ 60 \log\left(\frac{v}{12}\right)
+ 20 \log\left(\frac{l}{91.4}\right) \\
&+ d_f \cdot d_l + 3.0;
\end{aligned}
\end{equation}
where $L_{\mathrm{S},0}(f)$ is the reference spectrum of 'average vessel', such that

\begin{equation}
\label{eq:LS0_piecewise}
L_{\mathrm{S},0}(f)=
\begin{cases}
-10 \log_{10}\!\Bigg(
\begin{aligned}
&10^{-1.06 \log_{10}(f) - 14.34}\\
&\quad+\,10^{\,3.32 \log_{10}(f) - 21.425}
\end{aligned}
\Bigg), & f<500~\mathrm{Hz},\\[6pt]
173.2 - 18 \log_{10}(f), & f>500~\mathrm{Hz}.
\end{cases}
\end{equation}
The correction factors for frequency and length, $d_f$ and $d_l$, are calculated by

\begin{equation}
\label{eq:randi_dl}
d_l = \frac{l_{\mathrm{m}}^{1.15}}{995}
\end{equation}

\begin{equation}
\label{eq:randi_df}
d_f =
\begin{cases}
8.1~\text{dB}, & f \leq 28.4~\text{Hz} \\[6pt]
22.3~\text{dB} - 9.77 \log_{10}\!\left({f}\right) & 28.4~\text{Hz} < f \leq 191.6~\text{Hz} \\[6pt]
0~\text{dB}, & f > 191.6~\text{Hz}
\end{cases}
\end{equation}

\subsection{The JOMOPANS-ECHO Model}
The JOMOPANS-ECHO model~\cite{macgillivray2021reference}, utilizes the formulation of the RANDI 3.1 model as mentioned in~\eqref{Eq:RANDI}, albeit with a change to the 'average vessel' speed reference - instead of generic vessel speed of 12 knots, type-specific speed parameters ($V_c$) are used according to Table \ref{tab:app1}.

\begin{table}[h]
\centering
\footnotesize
\begin{tabular}{|l|l|c|}
\hline
\textbf{Vessel Class} & \textbf{AIS Type} & \textbf{$V_C$ (kn)} \\
\hline
(1) Fishing         & 30                           & 6.4  \\ \hline
(2) Tug             & 31, 32, 52                   & 3.7  \\ \hline
(3) Naval           & 35                           & 11.1 \\ \hline
(4) Recreational    & 36, 37                       & 10.6 \\ \hline
(5) Government/Research & 51, 53, 55              & 8.0  \\ \hline
(6) Cruise          & 60--69 (length $l > 100$ m)  & 17.1 \\ \hline
(7) Passenger       & 60--69 (length $l \leq 100$ m) & 9.7 \\ \hline
(8) Bulk          & 70, 75--79 (speed $V \leq 16$ kn) & 13.9 \\ \hline
(9) Container   & \makecell{71--74 (all speeds), 70, 75--79 \\ (speed $V > 16$ kn)} & 18.0 \\ \hline
(10) Vehicle Carrier & n/a                          & 15.8 \\ \hline
(11) Tanker          & 80--89                       & 12.4 \\ \hline
(12) Other           & All other type IDs           & 7.4  \\ \hline
(13) Dredger         & 33                           & 9.5  \\ \hline
\end{tabular}
\caption{Reference speeds ($V_C$) per vessel class in the JOMOPANS--ECHO model.}
\label{tab:app1}
\end{table}
The reference spectrum \(L_{\mathrm{S0}}(f)\) is defined by~\eqref{eq:LS0_base} for all vessel classes. For vessel classes 8--11, \eqref{eq:LS0_lowfreq_8to11} is implemented if \(f<100~\mathrm{Hz}\). The following formulations specify the values of the constants \(D\) and \(D^{\mathrm{LF}}\).

\begin{equation}
\label{eq:LS0_base}
\begin{aligned}
L_{\mathrm{S0}}(f)
&= 191
- 20\log_{10}\!\left(480\,\frac{v}{v_c}\right) \\
&\quad
- 10\log_{10}\!\left[
\left(1-\frac{f}{480\,\frac{v}{v_c}}\right)^{2}+D^{2}
\right].
\end{aligned}
\end{equation}

\begin{equation}
\label{eq:LS0_lowfreq_8to11}
\begin{aligned}
L_{\mathrm{S0}}(f)
&= 208
- 40\log_{10}\!\left(600\,\frac{v}{v_c}\right)
+ 10\log_{10}(f) \\
&\quad
- 10\log_{10}\!\left[
\left(1-\left(\frac{f}{600\,\frac{v}{v_c}}\right)^{2}\right)^{2}
+\left(D^{\mathrm{LF}}\right)^{2}
\right], \\
&\hspace{2.9em}
\left(f<100~\mathrm{Hz},\ \textrm{classes 8--11}\right).
\end{aligned}
\end{equation}

\begin{equation}
\label{eq:D_def}
D=
\begin{cases}
3, & \textrm{all vessel classes except 6},\\
4, & \textrm{vessel class 6}.
\end{cases}
\end{equation}

\begin{equation}
\label{eq:DLF_def}
D^{\mathrm{LF}}=
\begin{cases}
0.8, & \textrm{vessel classes 8--9},\\
1, & \textrm{vessel classes 10--11}.
\end{cases}
\end{equation}

\subsection{The LBDS Model}
The length, breadth, draft, and speed (LBDS) model~\cite{simard2016analysis} is constructed of polynomial terms, the sum of which provides the estimated $SL$ within a specified standard 1/3 octave band. The LBDS model is defined as

\begin{equation}
\label{eq:lbds_sl}
\begin{aligned}
\mathrm{SL}(f)_{1/3\text{ Octave}}
&= 285.40 + 0.0496\,f - 4.8\times 10^{-7}\,(f-2108.26)^2 \\[-1pt]
&\quad - 69.33\,\log_{10}(f) \\
&\quad - 49.29\bigl(\log_{10}(f)-2.70016\bigr)^2
      - 58.50\bigl(\log_{10}(f)-2.70016\bigr)^3 \\
&\quad - 41.54\bigl(\log_{10}(f)-2.70016\bigr)^4
      - 7.62\bigl(\log_{10}(f)-2.70016\bigr)^5 \\
&\quad + 13.47\,\log_{10}(l) - 0.55\,b + 0.0008\,(b-26.8854)^3 + 0.706\,d \\
&\quad + 20.164\,\log_{10}(v)
      - 505.1\bigl(\log_{10}(v)-1.12024\bigr)^3 \\
&\quad
      + 2891.9\bigl(\log_{10}(v)-1.12024\bigr)^5;
\end{aligned}
\end{equation}
where $f$ is the center frequency of a standard one-third octave band, $l$ is the length of the vessel in meters, $b$ is the breadth of the vessel in meters, $v$ is the speed of the vessel in knots, and $d$ is the draft of the vessel in meters.

\subsection{The AQUO Model}
AQUO~\cite{Audoly2015AQUO_R2_9} models the overall URN spectrum as the combination of dominant noise contributors: machinery noise, propeller non-cavitating noise, and cavitation noise. The combined level is determined by summing the component contributions in the power domain, as given as

\begin{equation}
\label{EQ:SLTOT}
\begin{aligned}
\textrm{SL}_{\textrm{TOT}}(f,V,L)
&=
10\log_{10}\!\Bigg[
10^{\frac{\textrm{SL}_{\textrm{mach}}(f,V,L_{\textrm{ref}})}{10}}
+
10^{\frac{\textrm{SL}_{\textrm{prop}}(f,V,L_{\textrm{ref}})}{10}} \\
&\hspace{3.0em}
+
10^{\frac{\textrm{SL}_{\textrm{cav}}(f,V,L_{\textrm{ref}})}{10}}
\Bigg]
+
25\log_{10}\!\left(\frac{L}{L_{\textrm{ref}}}\right).
\end{aligned}
\end{equation}
Where $\textrm{SL}_{\mathrm{TOT}}$ is the estimated power spectral density at frequency $f$, $\textrm{SL}_{\mathrm{mach}}$ is the noise contribution of machinery noise, $\textrm{SL}_{\mathrm{prop}}$ is the noise contribution of non-cavitating propeller noise, $\textrm{SL}_{\mathrm{cav}}$  is the noise contribution of cavitation noise, $V$ is the vessel speed (kn), and $L$ is the vessel length (m). The expression includes an additional logarithmic scaling term that accounts for the dependence on vessel size by relating $L$ to the class-specific reference length $L_{\mathrm{ref}}$.

The reference parameters and validity domains for vessel classes in the AQUO model are described in \ref{tab:aquo_categories} below.

\begin{table}[htbp]
\centering
\footnotesize
\setlength{\tabcolsep}{4pt} 
\renewcommand{\arraystretch}{1.15}
\begin{tabular}{|l|c|c|c|c|}
\hline
\textbf{\makecell{Vessel\\Class}} &
\textbf{\makecell{Ref.\\length\\(m)}} &
\textbf{\makecell{Ref.\\speed\\(kts)}} &
\textbf{\makecell{Length\\domain\\(m)}} &
\textbf{\makecell{Speed\\domain\\(kts)}} \\
\hline
High speed (large size)         & 150 & 30 & 80--200  & 10--40 \\ \hline
Cargo                           & 180 & 14 & 100--250 & 8--20  \\ \hline
Large cargo or container ship   & 280 & 20 & 250--350 & 10--25 \\ \hline
Tanker                          & 180 & 14 & 100--250 & 8--20  \\ \hline
Large tanker                    & 280 & 20 & 250--350 & 10--25 \\ \hline
Ferry                           & 180 & 18 & 100--250 & 8--25  \\ \hline
Large cruise vessel             & 250 & 18 & 200--300 & 8--25  \\ \hline
Fishing vessel                  & 50  & 10 & 40--70   & 6--12  \\ \hline
Research New                         & 80  & 11 & 50--100  & 6--12  \\ \hline
Research Old                         & 80  & 11 & 50--100  & 6--12  \\ \hline
Leisure craft (large size)      & 50  & 15 & 30--70   & 6--20  \\ \hline
Tug                             & 40  & 13 & 30--60   & 8--15  \\ \hline
Sailing boat (using engine)     & N/A & N/A & N/A     & N/A    \\ \hline
\end{tabular}
\caption{Reference parameters and validity domains for vessel classes in the AQUO model.}
\label{tab:aquo_categories}
\end{table}
To determine the parameters of machinery noise ($\textrm{SL}_{\mathrm{mach}}$), propeller non-cavitating noise ($\textrm{SL}_{\mathrm{prop}}$), and cavitation noise ($\textrm{SL}_{\mathrm{cav}}$), the AQUO model defines a frequency-dependent piecewise equation for each vessel class described in Table~\ref{tab:aquo_categories}. For example, for 'Cargo' vessels, the formulation is defined as

\begin{equation}
\label{eq:aquo_components}
\begin{alignedat}{2}
&L_{\mathrm{ref}}=180~\mathrm{m},\quad
V_{\mathrm{ref}}=14~\mathrm{kts},\quad
V_{\mathrm{cav}}=10~\mathrm{kts}.\\[6pt]
&\mathrm{SL}_{\mathrm{mach}}(f,V,L_{\mathrm{ref}})
=136+15\log_{10}V,
\qquad &&\text{for } f<200~\mathrm{Hz},\\
&\mathrm{SL}_{\mathrm{mach}}(f,V,L_{\mathrm{ref}})
=186-22\log_{10}f+15\log_{10}V,
&&\text{for } f\ge 200~\mathrm{Hz},\\[6pt]
&\mathrm{SL}_{\mathrm{prop}}(f,V,L_{\mathrm{ref}})
=109-5\log_{10}f+50\log_{10}V,
&&\text{for } f<80~\mathrm{Hz},\\
&\mathrm{SL}_{\mathrm{prop}}(f,V,L_{\mathrm{ref}})
=156-30\log_{10}f+50\log_{10}V,
&&\text{for } f\ge 80~\mathrm{Hz},\\[6pt]
&\mathrm{SL}_{\mathrm{cav}}(f,V,L_{\mathrm{ref}})
=79+10\log_{10}f+60\log_{10}V,
&&\text{for } f<50~\mathrm{Hz},\\
&&&\text{and } V>V_{\mathrm{cav}},\\
&\mathrm{SL}_{\mathrm{cav}}(f,V,L_{\mathrm{ref}})
=129-20\log_{10}f+60\log_{10}V,
&&\text{for } f\ge 50~\mathrm{Hz},\\
&&&\text{and } V>V_{\mathrm{cav}}.
\end{alignedat}
\end{equation}

\subsection{The SRV Model}
\label{sec:SRV}

The small recreational vessel (SRV) model~\cite{Shipton2026SmallVesselURN} provides a parametric representation of V-URN from small recreational vessels using a regularized log-linear regression formulation with Gaussian smoothing in log-frequency. Separate models are fitted for sailboats (under motor operation) and motorized yachts, yielding length- and speed-dependent source PSD estimates across the analyzed band. For each vessel class, a set of frequency-dependent regression coefficients is estimated and used to predict the source PSD level as a function of vessel length and speed.

The SRV model can be written as a vessel-class-specific source PSD predictor. For vessel class
$q \in \{\mathrm{sail},\mathrm{yacht}\}$, the predicted source PSD level at frequency $f$ is

\begin{equation}
\label{eq:srv_loglinear}
\textrm{SL}_{\mathrm{SRV},q}(f;L,V)
=
\beta_{0,q}(f)
+
\beta_{L,q}(f)\log_{10}(L)
+
\beta_{V,q}(f)\log_{10}(V),
\end{equation}
where $\textrm{SL}_{\mathrm{SRV},q}$ is expressed in dB re $1~\mu\mathrm{Pa}^{2}/\mathrm{Hz}$ at 1~m, $L$ is vessel length in m, $V$ is vessel speed in kn, and $\beta_{0,q}(f)$, $\beta_{L,q}(f)$, and $\beta_{V,q}(f)$ are frequency-dependent regression coefficients for vessel class $q$. The model is fitted independently for each frequency bin and for each vessel class, with L2 regularization used to stabilize the coefficient estimates.
Because the fitted coefficients may contain narrowband vessel-specific structure, the final SRV model uses log-frequency Gaussian smoothing. Defining $x=\log_2(f)$, the smoothed coefficient at frequency $f_i$ is

\begin{equation}
\label{eq:srv_smoothing}
\widetilde{\beta}_{j,q}(f_i)
=
\frac{
\sum_{k} 
\exp\!\left[
-\frac{\left(\log_2 f_k-\log_2 f_i\right)^2}{2\sigma^2}
\right]
\beta_{j,q}(f_k)
}{
\sum_{k}
\exp\!\left[
-\frac{\left(\log_2 f_k-\log_2 f_i\right)^2}{2\sigma^2}
\right]
},
\end{equation}
where $j \in \{0,L,V\}$, $f_k$ denotes the discrete frequency bins used for fitting, $\widetilde{\beta}_{j,q}$ are the smoothed regression coefficients, and $\sigma$ is the Gaussian width in log-frequency units, selected such that the kernel has a one-third-octave full width at half maximum. This yields the final SRV formulation as

\begin{equation}
\label{eq:srv_final}
\textrm{SL}_{\mathrm{SRV},q}(f;L,V)
=
\widetilde{\beta}_{0,q}(f)
+
\widetilde{\beta}_{L,q}(f)\log_{10}(L)
+
\widetilde{\beta}_{V,q}(f)\log_{10}(V).
\end{equation}

For use in one-third-octave or broadband noise mapping, the predicted PSD may be energetically integrated over a frequency band $B$ as

\begin{equation}
\label{eq:srv_band}
\textrm{SL}_{\mathrm{SRV},q}(B;L,V)
=
10\log_{10}
\left(
\int_{B}
10^{\textrm{SL}_{\mathrm{SRV},q}(f;L,V)/10}
\,df
\right),
\end{equation}
where $B$ is the frequency band over which the PSD prediction is energetically integrated.

\clearpage
\newpage

\bibliographystyle{elsarticle-harv} 
\bibliography{References}

\end{document}